\documentclass[twocolumn]{emulateapj}

\usepackage{graphicx}
\usepackage{amssymb}
\usepackage{threeparttable}
\usepackage{url}

\mathchardef\mhyphen="2D
\slugcomment{Submitted to MNRAS, Feb 18 2014}
\shorttitle{Quasar outflows}
\shortauthors{Zakamska \& Greene}

\begin{document}

\title{Quasar feedback and the origin of radio emission in radio-quiet quasars}

\author{Nadia L. Zakamska\altaffilmark{1}}
\author{Jenny E. Greene\altaffilmark{2}}
\altaffiltext{1}{Department of Physics \& Astronomy, Johns Hopkins University, Bloomberg Center, 3400 N. Charles St., Baltimore, MD 21218, USA}
\altaffiltext{2}{Department of Astrophysical Sciences, Princeton University, Princeton, NJ 08544, USA}

\begin{abstract}
We analyze SDSS spectra of 568 obscured luminous quasars. The [OIII]$\lambda$5007\AA\ emission line shows blueshifts and blue excess, indicating that some of the narrow-line gas is undergoing an organized outflow. The velocity width containing 90\% of line power ranges from 370 to 4780 km/sec, suggesting outflow velocities up to $\sim$2000 km/sec, and is strongly correlated with the radio luminosity among the radio-quiet quasars. We propose that radio emission in radio-quiet quasars is due to relativistic particles accelerated in the shocks within the quasar-driven outflows; star formation in quasar hosts is insufficient to explain the observed radio emission. The median radio luminosity of the sample of $\nu L_{\nu}$[1.4GHz]$=10^{40}$ erg/sec suggests a median kinetic luminosity of the quasar-driven wind of $L_{\rm wind}=3\times 10^{44}$ erg/sec, or about 4\% of the estimated median bolometric luminosity $L_{\rm bol}=8\times 10^{45}$ erg/sec. Furthermore, the velocity width of [OIII] is positively correlated with mid-infrared luminosity, which suggests that outflows are ultimately driven by the radiative output of the quasar. Emission lines characteristic of shocks in quasi-neutral medium increase with the velocity of the outflow, which we take as evidence of quasar-driven winds propagating into the interstellar medium of the host galaxy. Quasar feedback appears to operate above the threshold luminosity of $L_{\rm bol}\sim 3\times 10^{45}$ erg/sec.
\end{abstract}


\section{Introduction}
\label{sec:intro}

Black hole feedback -- the strong interaction between the energy output of supermassive black holes and their surrounding environments -- is routinely invoked to explain the absence of overly luminous galaxies, the black hole vs. bulge correlations and the similarity of black hole accretion and star formation histories \citep{tabo93, silk98, spri05, hopk06}. After years of intense observational effort, specific examples of black-hole-driven winds have now been identified using a variety of observational techniques, both at low and at high redshifts \citep{nesv06, arav08, nesv08, moe09, dunn10, alex10, harr12, fabi12}. 

How these outflows are launched near the black hole and established over the entire host galaxy remains a topic of active research. In particular, it is becoming clear that as the outflow impacts an inhomogeneous interstellar medium of the galaxy, the winds are expected to contain gas at a wide range of physical conditions, and these different ``phases'' of the winds require different types of observations \citep{veil05}. As a result, the determination of the physical parameters of these outflows -- including such basic parameters as the mass, the momentum and the kinetic energies they carry -- remains challenging. In the last few years, a lot of progress in this area has been made by observations of the coldest components of the outflows which are in the form of neutral or even molecular gas \citep{fisc10, feru10, rupk11, stur11, aalt12, veil13, rupk13a, rupk13b, cico14, sun14}. It remains unclear how common this cold component is in winds driven by a powerful active nucleus and how the mass, momentum and energy carried by the wind are distributed across the different phases. 

Several years ago, we embarked on an observational program to determine whether radio-quiet, luminous quasars have observable effects on their galaxy-wide environment. One of our lines of investigation is to determine the extent and kinematics of the warm ($T\sim 10^4$ K) ionized gas -- the so-called narrow-line region of quasars. We simplify the observational task by looking at obscured quasars \citep{zaka03, reye08} -- those where the line of sight to the nucleus is blocked by intervening material, allowing us to study the distribution of matter in the galaxy unimpeded by the bright central source. We find strong evidence that ionized gas is extended over scales comparable to or exceeding that of the host galaxy; furthermore, it is kinematically disturbed and is not in equilibrium with the gravitational potential of the galaxy \citep{gree09, gree11, hain13}. 

More recently, we surveyed a sample of obscured radio-quiet quasars using a spectroscopic integral field unit \citep{liu13a, liu13b}. We found extended ionized gas encompassing the entire host galaxy (median diameter of nebulae of 28 kpc), suggestive of wide-angle outflows, and determined kinetic energies of these outflows to be well in excess of $10^{44}$ erg/sec, with a median 2\% conversion rate from the bolometric luminosity to the kinetic energy of warm ionized gas; more energy can be carried by other components. Furthermore, we identified several candidate objects where the wind has ``broken out'' of the denser regions of the galaxy and is now expanding into the intergalactic medium, sometimes in bubble-like structures \citep{gree12, liu13b}. 

These observations demonstrate the presence of extended ionized gas in host galaxies of type 2 quasars, which is apparently out of dynamical equilibrium with the host galaxy and is likely in an outflow on the way out of the host. In this paper, we examine spectra of several hundred obscured quasars and we study the relationships between gas kinematics and other physical properties of these objects. In Section \ref{sec:data} we describe the sample selection, the dataset and the measurements. In Section \ref{sec:optical}, we conduct kinematic analysis of the optical emission lines. In Section \ref{sec:multiwv} we discuss the relationships between multi-wavelength properties of quasars and  kinematic measures of their ionized gas nebulae. In Section \ref{sec:compo} we present composite spectra and discuss trends in weak emission lines. We present qualitative models for radio emission and emission lines in Section \ref{sec:discussion}, and we summarize in Section \ref{sec:conclusions}. 

We use a $h$=0.7, $\Omega_m$=0.3, $\Omega_{\Lambda}$=0.7 cosmology throughout this paper. SDSS uses vacuum wavelengths, but for consistency with previous literature we use air wavelengths in Angstroms to designate emission lines. Wavelengths are obtained from NIST \citep{kram13} and Atomic Line List\footnote{\url{http://www.pa.uky.edu/~peter/atomic/}} and converted between air and vacuum as necessary using \citet{mort91}. Objects are identified as SDSS Jhhmm+ddmm, with full coordinates given in the catalog by \citet{reye08}. We use `1D', `2D' and `3D' abbreviations for one-, two-, and three-dimensional values. 

\section{Data and measurements}
\label{sec:data}

\subsection{Sample selection and host galaxy subtraction}
\label{sec:selection}

The obscured quasar candidates studied here were selected from the spectroscopic data of the Sloan Digital Sky Survey \citep{york00} based on their emission line ratios and widths to be the luminous analogs of Seyfert 2 galaxies \citep{zaka03}; the most recent sample contains 887 objects at $z<0.8$ \citep{reye08}. Infrared observations demonstrate that these sources have high bolometric luminosities (up to $10^{47}$ erg/sec, \citealt{zaka04, zaka08, liu09}). Chandra and XMM-Newton observations show that they contain luminous X-ray sources with large amounts of obscuration along the line of sight \citep{zaka04, ptak06, vign10, jia13}. HST imaging and ground-based spectropolarimetry demonstrate the presence of scattered light -- a classical signature of a buried broad-line active nucleus \citep{anto85, zaka05, zaka06}. In other words, all follow-up observations thus far are consistent with these objects being luminous obscured quasars. Our estimate of the number density of these objects suggests that they are at least as common as unobscured quasars at the same redshifts and line luminosities \citep{reye08}. 

\begin{figure}
\centering
\includegraphics[scale=0.7, clip=true, trim=0cm 10cm 10cm 0cm]{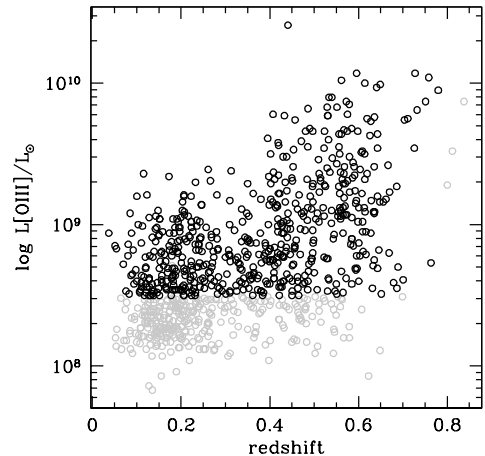}
\caption{The distribution of the entire \citet{reye08} sample of type 2 quasars in the redshift / [OIII] luminosity space (grey) and the 568 objects with kinematic analysis in this paper (black).}
\label{pic_sample}
\end{figure}

In this paper we examine the kinematic structure of narrow emission line gas in 568 objects (out of the entire sample of 887 by \citealt{reye08}) selected to have [OIII] luminosities above $10^{8.5}L_{\odot}$. Their distribution in [OIII] luminosity / redshift space is shown in Figure \ref{pic_sample}. At the median redshift of the sample presented here $z=0.397$, the SDSS fiber (3\arcsec\ in diameter) covers the galaxies out to 8 kpc away from the center. 

Ideally, we would like to measure the kinematics of the ionized gas relative to the host galaxy potential. The SDSS spectroscopic pipeline provides a high-quality redshift based on fits of observed spectra to a variety of library templates. In the cases of our objects, this pipeline latches onto the strong emission lines, so the redshifts are affected by narrow line kinematics and may be offset from the redshifts of the host galaxies. Thus, our first step is to determine the host galaxy redshifts based on the absorption features produced in stellar photospheres. 

Even though the quasars are obscured, the continuum from the host galaxy is very difficult to detect. One component of the continuum is due to the stars in the host galaxy. Furthermore, while the direct emission from the quasar is completely blocked in most cases, some quasar light reaches the observer after scattering off of the interstellar medium of the host galaxy and becomes an important continuum contribution at high quasar luminosities. Finally, in very luminous cases the Balmer continuum produced by free-bound transitions in the extended narrow emission line gas is also seen \citep{zaka05}, and the emission lines tend to confuse the search for stellar features in the continuum. 

We use the stellar velocity dispersion code described in detail in \citet{gree06} and \citet{gree09} to model the host galaxy continuum and to establish the systemic velocity. The continuum of each quasar is modeled as the linear combination of three stellar models plus a power-law component to mimic a possible scattered light contribution \citep{zaka06, liu09}. For templates, we use \citet{bruz03} single stellar population models rather than individual stellar spectra.

We first shift the spectra into the approximate rest-frame by using the SDSS pipeline redshifts. Then we fit the continuum over the wavelength range 3680-5450\AA, allowing for the stellar models to have a velocity of up to 300 km/sec relative to the SDSS frame and to be broadened with a Gaussian function that represents the stellar velocity dispersion of the host. When the host galaxy is well detected, typical stellar absorption features visible in the spectra include Ca H+K, G-band, and Mg I{\it b} lines. After this procedure, the entire host continuum is subtracted and the spectrum is shifted into the fine-tuned host galaxy frame. 

We are able to identify some host galaxy features in 271 objects. In most of these cases, the absorption features are so weak that we do not consider the reported host velocity dispersions to be accurate. The greater benefit of the host subtraction procedure is that in these 271 objects we can analyze narrow line kinematics relative to an accurately determined host frame. In the remaining objects we find no evidence for stellar features, so we subtract a featureless continuum. For the majority of objects, our workable wavelength range covers [OII]$\lambda\lambda$3726,3729, [OIII]$\lambda$5007 and everything in between. 

\subsection{Fitting functions and non-parametric measurements}

We aim to use non-parametric measures that do not depend strongly on the specific fitting procedure. We need robust measures or robust analogs of the first four moments of the line profile: typical average velocity, velocity dispersion, and the skewness and the kurtosis of the velocity distribution. We fit the profiles with one to three Gaussian components in velocity space, but in principle other fitting functions could be used. We assign no particular physical significance to any of the parameters of the individual components; rather, the goal is to obtain a noiseless approximation to the velocity profile. 

\begin{figure}
\centering
\includegraphics[scale=0.45]{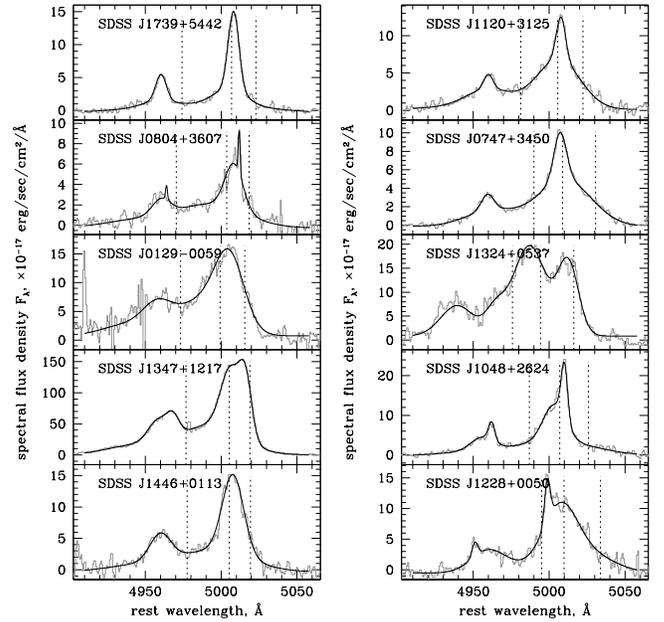}
\caption{Spectra of the [OIII]$\lambda\lambda$4959,5007 doublet in the ten objects with the highest $w_{80}$ values ($2314\le w_{80}\le 2918$ km/sec), with their multi-Gaussian fits. The two lines in the doublet are fit simultaneously, under the assumption that the kinematic structure of both lines is the same and that the ratio of amplitudes is 0.337. Dashed lines show the positions of $v_{10}$, $v_{50}$ and $v_{90}$. }
\label{pic_example}
\end{figure}

We use relative change in reduced $\chi^2$ values to evaluate which fit should be accepted; if adding an extra Gaussian component leads to a decrease in $\chi^2$ of $<$10\%, we accept the fit with a smaller number of components. The single-Gaussian fit is accepted for 36 objects, a two-Gaussian fit is accepted for 132 and the remaining 400 objects are fit with three Gaussians. Almost all objects that have high signal-to-noise observations have reduced $\chi^2$ values that are too high to be statistically acceptable and thus would require either a larger number of components or different fitting functions to be fitted to statistical perfection. Fortunately, the non-parametric measures that we derive are rather robust: adding the third Gaussian component changes our second moment measure $w_{80}$ by less than 10\% in 83\% of objects. Examples of line fits are shown in Figure \ref{pic_example}. The objects selected for this figure have the top ten highest values of $w_{80}$ (our analog of the velocity dispersion defined below). 

Armed with fitting functions performed in velocity space $f(v)$, we construct the normalized cumulative velocity distribution $F(v)=\int_{-\infty}^vf(v'){\rm d}v'/\int_{-\infty}^{+\infty}f(v'){\rm d}v'$. Since the velocity profile is a noiseless non-negative function, $F(v)$ is strictly monotonically increasing. We then determine the velocities at which 5\%, 25\%, 50\%, 75\% and 95\% of the line flux accumulates. The median velocity $v_{50}$ is the solution of the equation $F(v)=0.5$. The width comprising 90\% of the flux is $w_{90}=v_{95}-v_{05}$, the width at 80\% is $w_{80}=v_{90}-v_{10}$ and the width comprising 50\% of the flux is $w_{50}=v_{75}-v_{25}$. All these values have dimensions of velocity (km/sec). For a Gaussian profile, the value $w_{80}$ is close to the conventionally used full width at half maximum ($w_{80}=2.563\sigma=1.088$FWHM; $w_{90}=3.290\sigma$). For a typical object in our sample (median $w_{80}=752$ km/sec) the instrumental dispersion of the SDSS ($\sigma_{\rm inst}=70$ km/sec) contributes only a few per cent to the line width \citep{liu13b}, and as the line profiles are typically non-Gaussian, we do not attempt to deconvolve the resolution except a couple of cases noted explicitly when we use Gaussian components individually. 

We can measure the asymmetry of the velocity profile relative to the median velocity by computing a dimensionless relative asymmetry $R=((v_{95}-v_{50})-(v_{50}-v_{05}))/(v_{95}-v_{05})$. Negative values correspond to cases where the blueshifted wing of the line extends to higher velocities than the redshifted one, and positive values correspond to cases where the redshifted wing dominates. This measure is a non-parametric analog of skewness and is equal to 0 for any symmetric profile (including a single Gaussian). Furthermore, we can measure the prominence of the wings of the profile, or the non-parametric analog of the kurtosis, by computing $r_{9050}\equiv w_{90}/w_{50}$. For a Gaussian profile, this value is equal to 2.4389. Values higher than this indicate profiles with relatively more extended wings than a Gaussian function: for example, a Lorentzian profile $f(v)=1/(\gamma^2+v^2)$ (where $\gamma$ is the measure of the profile width) has $r_{9050}=6.3138$. Values lower than the Gaussian value indicate a profile with a stronger peak-to-wings ratio and are rarely encountered in our sample. 

Finally, we compute the absolute asymmetry of the profile, which is $A=$(flux($v>0$)-flux($v<0$))/total flux. In terms of the normalized cumulative velocity distribution, $A=1-2F(0)$. This asymmetry is dimensionless and it is positive for profiles with more flux at redshifted wavelengths than at blueshifted wavelengths. 

Values $A$ and $v_{50}$ critically depend on an accurate determination of the host galaxy redshift, because this is what we use to fix the $v=0$ point. If no absorption features in the composite stellar light of the host are detected, the redshift can only be determined from the emission lines themselves, which renders the absolute velocity and skewness meaningless. Values of $R$, $w_{90}$, $w_{80}$, $w_{50}$, and $r_{9050}$ include only differences between velocities and do not hinge on the accurate determination of the host velocity. 

\subsection{Robustness of non-parametric measures}
\label{sec:robust}

In this section we evaluate the performance of the non-parametric measures. The theoretical advantage of the non-parametric measures is in their relative insensitivity to the fitting functions used. We test this assumption by repeating all the fits using a set of one, two or three Lorentzian ($f(v)=1/(\gamma^2+v^2)$) profiles. The Lorentzian function has significantly more flux in the faint wings than does the Gaussian function, and this shape is not borne out in the observations of the line profiles in our sample. The quality of the fits with Lorentzian profiles is significantly poorer than of those with Gaussian profiles, and therefore all our final non-parametric measures are based on multi-Gaussian fitting as described in the previous subsection. 

Nevertheless, we carry out the comparison between the non-parametric measures derived from the two methods. We find that both sets of fits yield nearly identical absolute asymmetries and median velocities $v_{50}$, which is not surprising because both these measures are most sensitive to the correct identification of the line centroid. All other measures (the widths, relative asymmetry and $r_{9050}$) are strongly correlated between the two sets of fits, but the specific values are systematically different. The line width $w_{80}$ as measured from the Lorentzian fits is about 25\% higher than that from the Gaussian fits; the relative asymmetry is significantly weaker as measured by Lorentzian profiles than the one measured by the Gaussian ones; and $r_{9050}$(Lorentzian) is approximately equal to $r_{9050}$(Gaussian)$+2$. All these differences are as expected from the fitting functions with different amounts of power in the extended wings. 

The conclusions we derive from this comparison are two-fold. First, since the multi-Lorentzian fits are not only statistically but also visibly inferior to the multi-Gaussian ones, the real systematic uncertainty on the $w_{80,90}$ -- the key measurements discussed in this paper -- is significantly smaller than the 25\% difference between line widths calculated from these two methods. This is very encouraging. (For the majority of objects, $w_{80}$ is accurate to 10\% or better, as measured from the comparison of non-parametric measures derived from two-Gaussian and three-Gaussian fits.) Second, we confirm that the non-parametric measures are relatively robust: although the Lorentzian profiles do not yield statistically good fits, they nevertheless give reasonable estimates of the non-parametric measures. 

We perform an additional test to determine the effect of the signal-to-noise ratio (S/N) of the spectra on our measurements. A narrow Gaussian emission line with a weak broad base observed with a high S/N is represented by two Gaussians in our multi-Gaussian fit, and its non-parametric measures include the power contributed by the weak broad base. On the contrary, if the same object is observed in a lower quality observation, the weak base is not necessarily recognized as such because the $\chi^2$ of the two-Gaussian fit may be indistinguishable from the one-Gaussian one and thus the latter will be preferred. 

\begin{figure*}
\centering
\includegraphics[scale=0.95, clip=true, trim=0cm 15.5cm 0cm 0cm]{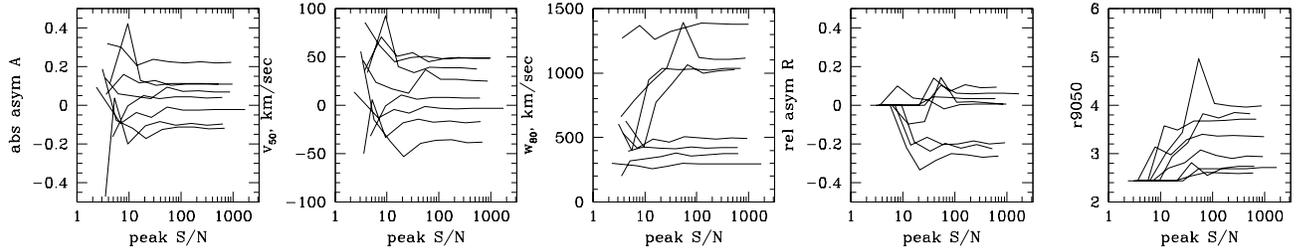}
\caption{The dependence of non-parametric measures on the peak signal-to-noise ratio of the [OIII] emission line for eight spectra whose quality is progressively artificially downgraded by adding Gaussian random noise. }
\label{pic_noise}
\end{figure*}

We conduct the following Monte Carlo test to explore the effect of noise on our measurements. We take eight of the highest S/N objects in our catalog, four with $w_{80}>1000$ km/sec and four with $w_{80}<500$ km/sec. We then downgrade the quality of these spectra by adding progressively higher Gaussian random noise to the original (essentially noiseless) spectra and conduct all our multi-Gaussian and non-parametric measures in the manner identical to that used for real science observations. The results are shown in Figure \ref{pic_noise}. Both absolute asymmetry and $v_{50}$ are relatively insensitive to the noise and do not show systematic trends. The line width $w_{80}$ does not depend on the noise for the four objects with relatively narrow ($w_{80}<500$ km/sec) lines. But for the broad-line objects the measured $w_{80}$ noticeably declines as the quality of observations worsens, reflecting the `missing broad base' phenomenon. Measurements with peak S/N$>10$ are relatively safe from this phenomenon: only one of the eight objects shows a noticeable decline of $w_{80}$ at S/N$\la 30$. The relative asymmetry and the kurtosis-like $r_{9050}$ quickly drop to single-Gaussian values as the S/N decreases below 20 or so. 

\section{Kinematic analysis of integrated spectra}
\label{sec:optical}

\subsection{Outflow signatures}
\label{sec:outflows}

Interpreting the line-of-sight gas kinematic measurements in terms of physical 3D motions of the gas is highly non-trivial. This is true even when spatial information is available, for example, via integral-field observations \citep{liu13b}, but is even more so when we have to rely only on one spatially integrated spectrum. The reason is that if an object exhibits a spherically symmetric optically thin outflow, then its emission line profiles are symmetric and peaked at zero velocity, since there is a large amount of gas moving close to the plane of the sky. Therefore, there is simply no ``smoking gun'' outflow signature in the emission line profile of such a source. If an outflow has non-zero optical depth, then its presence can be inferred by its absorption of the background light at wavelengths blueshifted relative to the quasar rest-frame, as happens in quasar absorption-line systems \citep{cren03, arav08}, but this is not our case. Therefore, proving that a given line profile is due to gaseous outflow is often difficult and relies on indirect arguments \citep{liu13b}. 

To investigate the relationship between the observed line-of-sight velocity dispersion of the lines and the typical outflow velocity, we consider for the moment a spherically symmetric outflow with a constant radial velocity $v_0$. Because different streamlines have different inclinations to the line of sight, the observer sees a range of velocities -- in this simplest case of the spatially integrated spectrum, the line profile is a top-hat between $-v_0$ and $v_0$, so that $w_{80}=1.6v_0$. In \citet{liu13b}, we consider an outflow with a power-law luminosity density and calculate the velocity profiles in a spatially resolved observation, finding a typical $w_{80}\simeq 1.3v_0$ in the outer parts of the narrow line region. If a flow consists of clouds moving on average radially with velocity $v_0$, but also having an isotropic velocity dispersion $\sigma$, then the observed $w_{80}$ can be approximately calculated as $w_{80}\simeq \sqrt{(1.6v_0)^2+(2.42\sigma)^2}$, as long as $v_0$ and $\sigma$ are not too dissimilar. 

Another simple case is a radial flow, in which at every point there are clouds with a range of radial velocities. As an example, we consider the results of the 2D simulations of quasar feedback by \citet{nova11}, in which at every distance from the quasar the higher density regions (`clouds') have a wide distribution of radial velocities (G.Novak, private communication) which ranges from 0 to $\ga 1000$ km/sec, with a median among all clouds of $v_0=220$ km/sec. We use the velocities of clouds straight from these simulations and several different luminosity density profiles to produce mock emission line profiles. 

\begin{figure}
\centering
\includegraphics[scale=0.6, clip=true, trim=0cm 10cm 10cm 0cm]{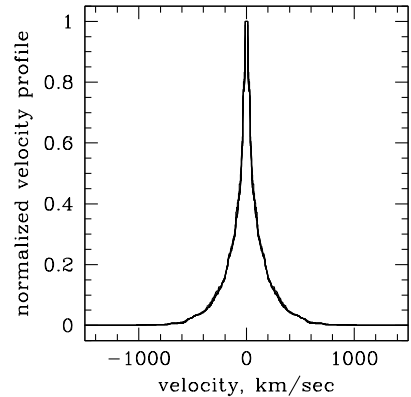}
\caption{Mock emission line velocity profiles constructed from the cloud velocity distribution from \citet{nova11}, using six different emissivity profiles (from linearly declining, to Gaussian, to flat, to centrally-tapered power law). All six curves are essentially on top of one another because the velocity distribution of the clouds hardly varies with the distance from the quasar in these simulations. The actual median radial velocity of the clouds is 220 km/sec, and we measure $w_{80}=310$ km/sec and $w_{90}=470$ km/sec for the simulated profile. }
\label{pic_novak}
\end{figure}

The resulting profiles are shown in Figure \ref{pic_novak}; they are insensitive to the adopted luminosity density profile because the velocity distribution of clouds does not change appreciably as a function of distance in these simulations. The profiles are peaked at zero velocity because there is a large population of clouds with small radial velocities, while the velocity dispersion of clouds is neglected in our calculations. If the clouds have an isotropic velocity dispersion which does not vary with distance, a more faithful profile can be obtained by convolving the profiles in Figure \ref{pic_novak} with a Gaussian, which would make the profiles broader and less peaky. The measured velocity width of the profile is $w_{80}\simeq 1.4 v_0$, similar to the scaling obtained in other simple cases, despite the broadness of the velocity distribution in this example. 

To sum up, (i) there is no tell-tale outflow signature in an optically thin, spherically symmetric outflow; (ii) deviations from spherical symmetry (and moreover from axial symmetry) are required to produce asymmetric line profiles; and (iii) the velocity width of the emission line can be used to estimate the outflow velocity, $w_{80}\simeq (1.4-1.6)\times v_0$. The most natural way in which the symmetry may be expected to be broken is due to dust obscuration, either by dust embedded in the outflow itself or by dust concentrated in the galactic disk. In either case, the redshifted part is more affected by extinction, and thus excess blueshifted emission is considered a sufficient indicator of an outflow \citep{heck81, dero84, whit85a, wils85}. In such case, the apparent $w_{80}$ decreases typically by $\la 30\%$ if the extinction is $\la 2.5$ mag and concentrated in a disk \citep{liu13b}, reducing $w_{80}/v_0$ by the same amount. Thus for a given $w_{80}$, an asymmetric profile indicates a somewhat higher $v_0$ than a symmetric one. 

Double-peaked profiles are expected in some geometries for a bi-conical outflow or more complex outflow kinematics \citep{cren00}, but can also be due to the rotation of the galaxy disk or two (or more) active nuclei in a merging system of galaxies, each illuminating its own narrow-line region. Distinguishing these possibilities usually requires follow-up observations at high spatial resolution, and the relative frequency of these scenarios remains a matter of debate \citep{come09a, liu10a, shen11, fu12, barr13, blec13}, but it appears that outflows dominate over dual active nuclei. It is likely that complex outflow kinematics is responsible for the majority of split-line profiles in our sample, and we conduct non-parametric kinematic measurements of such objects in the same way we do for the rest of the sample and include them in all our analyses.

\subsection{Analysis of [OIII] kinematics}
\label{sec:analysis}

\begin{figure*}
\centering
\includegraphics[scale=0.95, clip=true, trim=0cm 1cm 0cm 0cm]{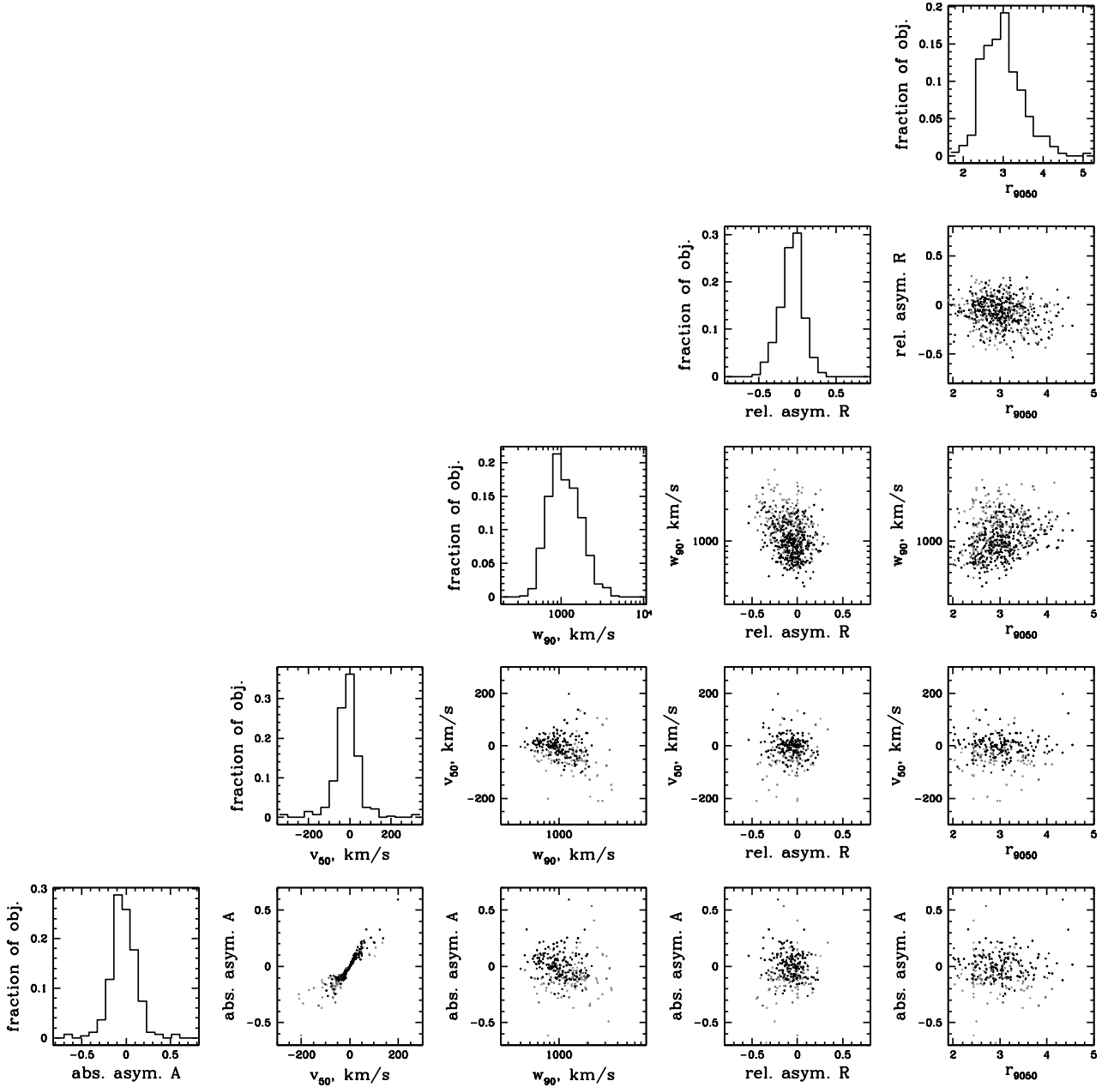}
\caption{Distributions of non-parametric kinematic measures and their mutual relationships. Points are color-coded from light grey to dark grey by the peak signal-to-noise ratios (lightest have S/N$\la 20$, grey have S/N from 20 to 50 and black have peak S/N$\ga 50$). In all the plots involving absolute asymmetry $A$ and median velocity $v_{50}$ only points with reliable host redshifts are displayed since both these values critically depend on the host redshift determination. }
\label{pic_np}
\end{figure*}

In Figure \ref{pic_np} we present the results of the non-parametric measurements of the [OIII] line in our sample of type 2 quasars. Both the relative asymmetry $R$ and the absolute asymmetry $A$ demonstrate slight preference for negative values, i.e., for blue excess. The blue asymmetries indicate that there is at least some outflow component in the [OIII]-emitting gas in type 2 quasars. The sample means and standard deviations for these values are $A=-0.03 \pm 0.15$ and $R=-0.08\pm 0.15$. Line widths, relative asymmetries, kurtosis parameters and (to a lesser extent) median velocities are correlated with one another, in the sense that objects with broader lines also have a more pronounced blue excess (negative $R$), higher $r_{9050}$ and more negative $v_{50}$ (for the latter, the sample mean and standard deviation are $-14\pm 75$ km/sec). Thus, as we discussed in the previous section, it is plausible that the [OIII] velocity width can serve as a proxy for the outflow velocity. Similar relationships have been reported by other authors in type 1 quasars, e.g., by \citet{stei13}, who find a stronger blueshift of [OIII] relative to broad H$\beta$ and [OII]$\lambda\lambda$3726,3729 as a function of [OIII] width.

\begin{figure}
\centering
\includegraphics[scale=0.45, clip=true, trim=0cm 10cm 0cm 0cm]{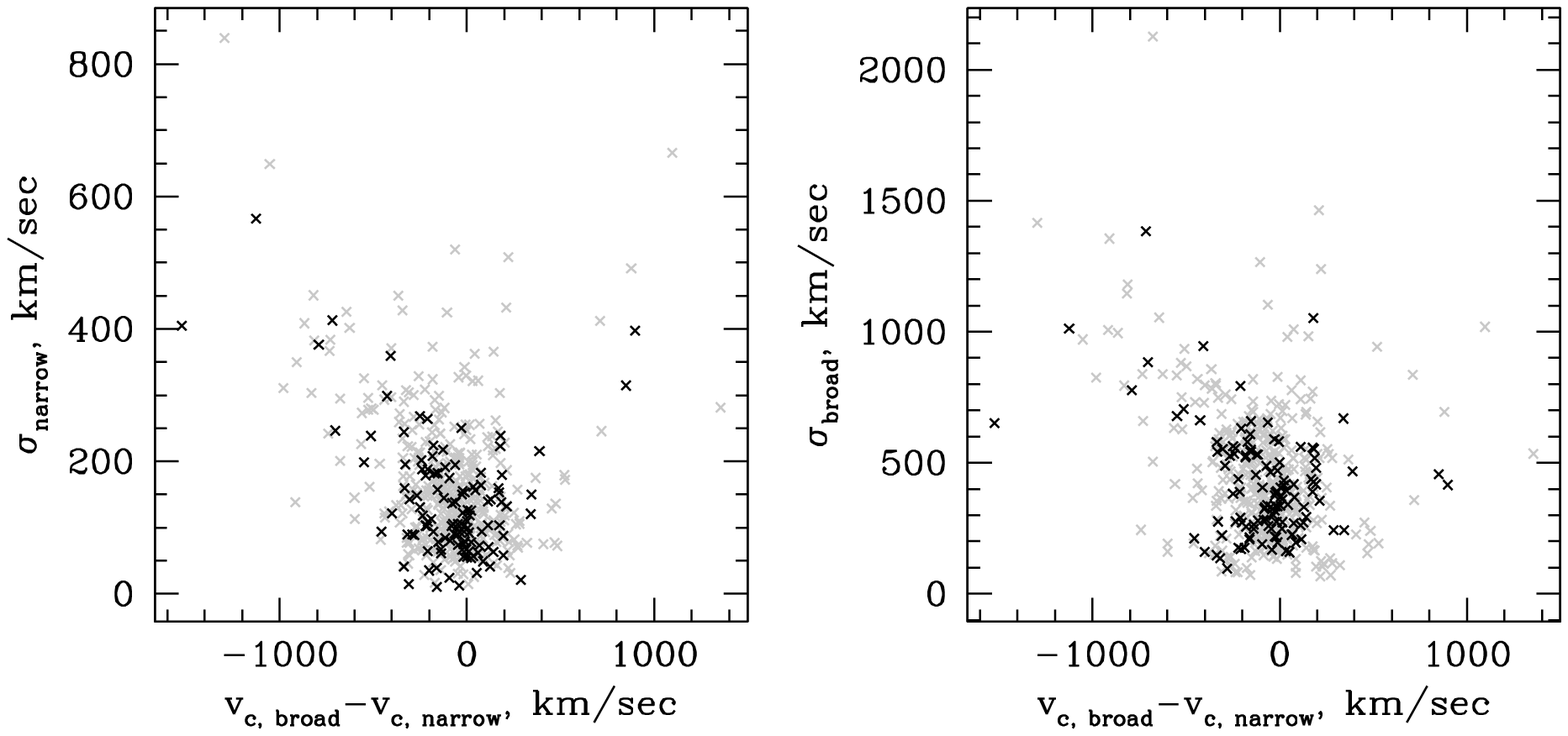}
\caption{Velocity offset between the centroids of the broad component from the narrow component (black for profiles decomposed into two Gaussians, grey for the two Gaussians that dominate the flux in three-Gaussian decompositions). $v_c$ is for the velocity centroids. Broad components tend to be blueshifted relative to the narrow ones. The velocity dispersions of individual Gaussian components have been corrected for the instrumental resolution ($\sigma_{\rm inst}=70$ km/sec subtracted in quadrature).}
\label{pic_vel}
\end{figure}

From these correlations, a picture emerges in which the outflow component, or at least the component of the outflow which is more affected by the obscuration, tends to be broad. We further explore this notion in Figure \ref{pic_vel}, where we split the line profiles into a `broad' and a `narrow' component. For objects with 2-Gaussian fits, the designation is straightforward, but the majority of objects require three Gaussians, in which case we pick the two most luminous ones and designate them `broad' and `narrow' according to their velocity dispersions. In Figure \ref{pic_vel}, we show that indeed the broader of the Gaussian components is the one that tends to be blueshifted relative to the narrow ones. The narrow cores tend to be well-centered in the host galaxy frame; the mean and standard deviation of the centroids of the narrower Gaussian components is $3\pm 150$ km/sec for the 271 objects with accurately determined host redshifts. On the contrary, the broader components tend to be slightly blueshifted, both relative to the host galaxy frame (velocity centroid of $-60\pm 210$ km/sec in the 271 objects with accurate host redshifts) and relative to the narrow components ($v_{\rm c, broad}-v_{\rm c, narrow}=-90 \pm 270$ km/sec for the entire sample of 568 sources). 

Because the narrow component is well-centered in the host frame, it is tempting to postulate that the narrow Gaussian component tends to be produced by gas in dynamical equilibrium with the host galaxy, e.g., in rotation in the galaxy disk, and is simply illuminated by the quasar, whereas the broad component is due to the outflow. However, we hesitate to make this inference, as there is no particular reason to assign any physical meaning to the individual parameters of the Gaussian components. We again draw a lesson here from Figure \ref{pic_novak}, where the mock emission line profile is due entirely to the outflowing clouds and can be decomposed into several Gaussian components, none of which correspond to the gas in rotation in the host galaxy. 

\begin{figure}
\centering
\includegraphics[scale=0.45, clip=true, trim=0cm 10cm 0cm 0cm]{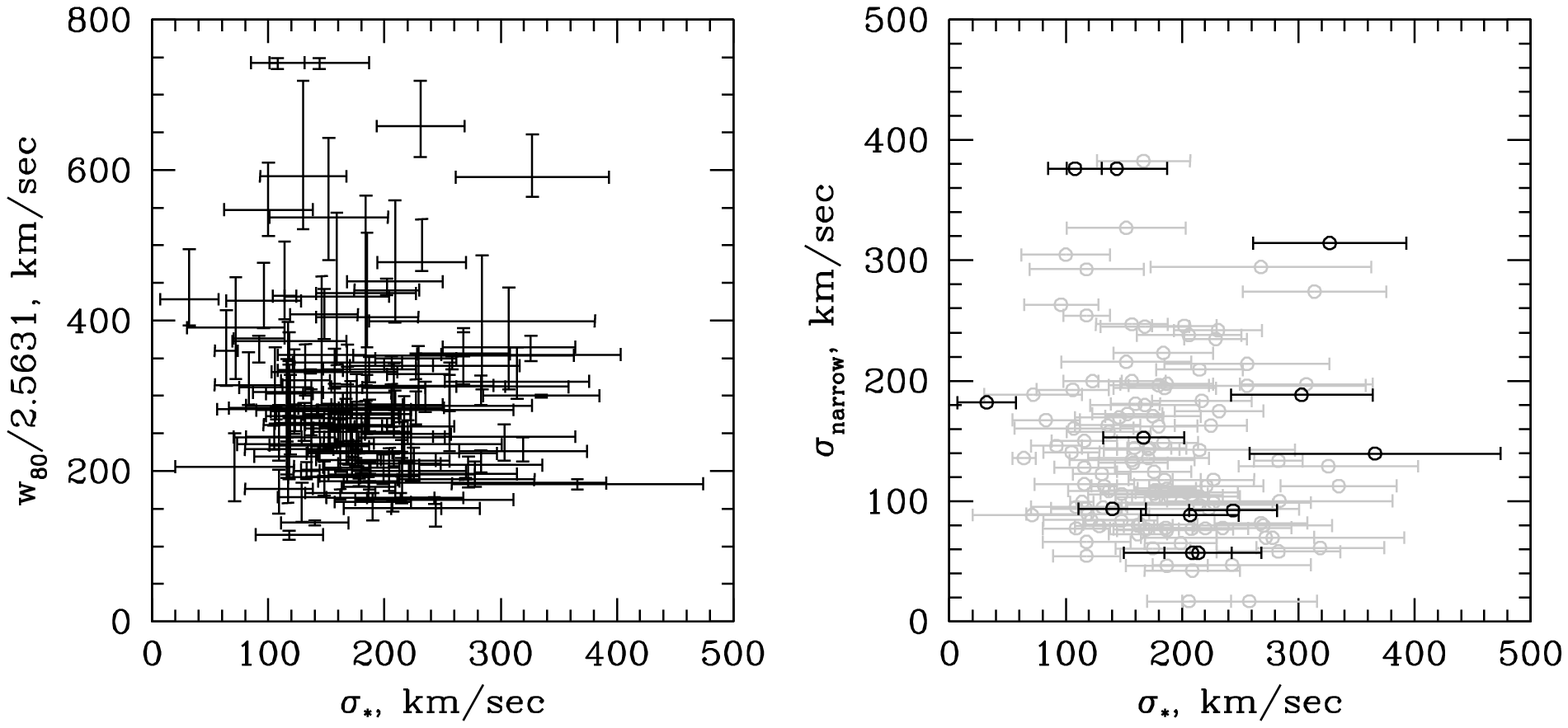}
\caption{Taking just the objects with well-determined stellar velocity dispersions from \citet{gree09}, we plot the overall width of the [OIII] emission line in the left panel and the dispersion of just the narrow component in the right panel as a function of the stellar velocity dispersion (grey points for the narrower of the two dominant components in a three-Gaussian fit, black points for the narrower of the two components in a two-Gaussian fit). The left panel is similar to Fig 7 of \citet{gree09} even though the exact non-parametric measures of dispersion used in that paper were defined differently. The error bars on $w_{80}$ reflect non-Gaussianity of lines and are calculated by converting from $w_{50}$ and $w_{90}$ assuming a Gaussian profile; thus, for a Gaussian profile they would be zero. Stellar velocity dispersions and $\sigma_{\rm narrow}$ have been corrected for instrumental dispersion; since lines are non-Gaussian, $w_{80}$ values are plotted as observed (typical correction is a few per cent).}
 \label{pic_stellar}
\end{figure}

We go further in Figure \ref{pic_stellar}, where we show that neither the overall line width nor the width of the narrower Gaussian component show any correlation with the stellar velocity dispersion. Since even among the 271 objects where the host galaxy was detected many of the stellar velocity dispersions are rather poorly determined for the reasons discussed in Section \ref{sec:selection}, for this figure we use only the better determined stellar velocity dispersions from \citet{gree09}. That the overall line width (left panel) shows no relationship with stellar velocity dispersion is not surprising if most of the line width arises due to the outflow. But the narrow cores do not appear to show any relationship with the stellar velocity dispersion either. As a result, it seems likely to us that virtually none of the [OIII]-emitting gas in type 2 quasars is in dynamical equilibrium with the host galaxy. This is in contrast to the situation in lower luminosity active galaxies in which [OIII] width strongly correlates with galaxy rotation and / or bulge velocity dispersion \citep{wils85, whit92b, nels96, gree05o3}. In such objects, it appears that the gas motions are in accord with the gravitational forces in the galaxy, and the gas is simply illuminated and photo-ionized by the active nucleus to produce the narrow-line region. 

The relative shifts between the narrower and the broader components could arise if the narrow component is produced on all scales in the host galaxy, where it is less likely to suffer from strong extinction, whereas the broader component is produced closer to the nucleus, where it is more likely to be affected by extinction. This is consistent with an outflow that is driven close to the nucleus and then gradually slowed down by the interactions with the interstellar medium \citep{wagn13}. Furthermore, this picture is consistent with the apparent decline of the line width in the outer parts of the outflow seen in the integral field unit observations of type 2 quasars \citep{liu13b}, although the effect is small, with velocity width declining only by 3\% per projected kpc. 

The sample mean and standard deviation of line width is $w_{90}=1230\pm 590$ km/sec, with a median of 1060 km/sec, minimum of 370 km/sec and maximum of 4780 km/sec ($w_{80}=880\pm 430$ km/sec, median 752 km/sec, min 280 km/sec, max 2918 km/sec), much higher than that of local ultraluminous infrared galaxies (ULIRGs; median $w_{90}\simeq 800$ km/sec) and especially of those without a powerful active nucleus in their center (median $w_{90}\simeq 600$ km/sec for pure starbursts, \citealt{hill14}). For ULIRGs, the line widths and the outflow velocities strongly correlate with the power source (higher for active galaxies, lower for starbursts; \citealt{rupk13a, hill14}), and the majority of the objects in our sample show line widths consistent with quasar-driven outflows, as expected. In \citet{liu13b}, we estimated that for gas disks rotating in the potential of the most massive galaxies line widths do not exceed $w_{80}\simeq 600$ km/sec. Thus, the line-of-sight gas velocities that we see in our sample are too high to be confined even by the most massive galaxy potential, and this gas cannot be in dynamical equilibrium with the host galaxy. 

The range of [OIII] luminosities in our sample is not all that large, with 90\% of sources between $\log(L{\rm [OIII]}/L_{\odot})=8.5$ and 9.5, and the remaining 10\% sources spanning the higher decade in luminosity. There is some tendency of objects with more pronounced outflow signatures (higher width, higher $r_{9050}$) to have higher [OIII] luminosity (Spearman rank correlation coefficient $r_{\rm S}\simeq 0.19$ for both relationships, probability of the null hypothesis of uncorrelated datasets $P_{\rm NH}=10^{-5}$), however, no correlations are seen between $L$[OIII] and absolute asymmetry, relative asymmetry or median velocity (Table \ref{tab:pnh}). Furthermore, any correlations between $L$[OIII] and kinematic measures can be strongly affected by the cutoff in the [OIII] luminosity distribution which is due to our sample selection ($\log(L{\rm [OIII]}/L_{\odot})\ge 8.5$. We report the significance of correlations at face value, without trying to account for the effects of the luminosity cutoff.

\begin{table*}
\caption{Relationships between gas kinematics and luminosity indicators}
\setlength{\tabcolsep}{1.4mm}
\label{tab:pnh}
\begin{center}
\begin{tiny}
\begin{tabular}{ | p{3cm} | p{2.1cm} | p{2.1cm} | p{2.1cm} | p{2.1cm} | p{2.1cm} | p{2.1cm} | }
\hline
Luminosity indicator & abs. asym. $A$ & med. vel. $v_{50}$ & width $w_{50}$ & width $w_{90}$ & rel. asym. $R$ & shape par. $r_{9050}$ \\
\hline
$\nu L_\nu$[1.4 GHz], upper limits at face value & 271; -0.37; $<10^{-5}$ & 271; -0.35; $<10^{-5}$ & 568; 0.30; $<10^{-5}$ & 568; 0.29; $<10^{-5}$ & 568; 0.06; 0.18 & 568; -0.01; 0.88 \\
$\nu L_\nu$[1.4 GHz], upper limits decreased by 2 & 271; -0.36; $<10^{-5}$ & 271; -0.35; $<10^{-5}$ & 568; 0.34; $<10^{-5}$ & 568; 0.34; $<10^{-5}$ & 568; 0.04; 0.30 & 568; 0.02; 0.72 \\
$L$[OIII] & 271; -0.05; 0.42 & 271; -0.03; 0.65 & 568; 0.12; $3\times 10^{-3}$ & 568; 0.19; $10^{-5}$ & 568; 0.06; 0.18 & 568; 0.19; $10^{-5}$ \\
$\nu L_\nu$[5\micron]  & 270; -0.15; 0.02 & 270; -0.16; $7\times 10^{-3}$ & 562; 0.36; $<10^{-5}$ & 562; 0.37; $<10^{-5}$ & 562; -0.11; 0.01 & 562; 0.04; 0.35 \\
$\nu L_\nu$[12\micron]  & 259; -0.18; $4\times 10^{-3}$ & 259; -0.19; $2\times 10^{-3}$; $v_{50}<0$: 151; -0.33; $4\times 10^{-5}$; $v_{50}\ge 0$: 108; 0.19; 0.04 & 539; 0.41; $<10^{-5}$ & 539; 0.44; $<10^{-5}$ & 539; -0.10; 0.02 & 539; 0.10; 0.02 \\
mid-infrared slope $\beta$ (higher is redder) & 259; -0.03; 0.59 & 259; -0.04; 0.54 & 539; 0.16; 3$\times 10^{-4}$ & 539; 0.23; $<10^{-5}$ & 539; -0.02; 0.69 & 539; 0.21; $<10^{-5}$ \\
$L_{\rm X}$, all available objects  & 24; 0.05; 0.82 & 24; 0.12; 0.56 & 54; 0.14; 0.32 & 54; 0.11; 0.43 & 54; 0.08; 0.56 & 54; 0.01; 0.93 \\
$L_{\rm X}$, Compton-thin objects  & 12; -0.29; 0.36 & 12; -0.13; 0.68 & 30; -0.05; 0.78 & 30; -0.05; 0.79 & 30; 0.38; 0.04 & 30; -0.31; 0.09 \\
\hline
\end{tabular}
\end{tiny}
\tablenotes{{\bf Notes.} Statistical relationships between luminosity indicators (listed in different rows) and kinematic indicators (listed in different columns). For every pair of variables, we give three values: the number of objects included, the Spearman correlation coefficient $r_{\rm S}$ and the significance $P_{\rm NH}$. Small values $P_{\rm NH}<0.01$ indicate a presence of a relationship (positive or negative correlation) at $>99$\% confidence level. The full sample has 568 objects; the subsample with accurate host velocities has 291 objects (a small number in each of these categories are missing reliable mid-infrared data because of blending); the subsample with X-ray detections has 54 objects; and the subsample of Compton-thin objects has 30 objects. $L_{\rm X}$ is the absorption-corrected intrinsic 2-10 keV luminosity from \citet{jia13}, and radio and infrared luminosities are corrected to the rest-frame. The significance of correlations with $L$[OIII] is quoted at face value even though the distribution of $L$[OIII] is artificially cut off by our target selection. We show a small subsample of the strongest correlations in the figures. }
\end{center}
\end{table*}

\subsection{Fainter lines}
\label{sec:shapes}

We compare the kinematic structure of the brightest lines, [OIII]$\lambda$5007, [OII]$\lambda\lambda$3726,3729, and H$\beta$. It has long been known that line kinematics often vary as a function of the ion ionization potential or the line critical density \citep{whit85c}. On the basis of largely anecdotal evidence, we previously established that in type 2 quasars with highly asymmetric or split-line [OIII]$\lambda$5007 profiles the [OII]$\lambda\lambda$3726,3729 profiles seemed less complex \citep{zaka03}. The difficulty of this analysis is illustrated in Figure \ref{pic_noise}: because [OIII]$\lambda$5007 is by far the brightest line, it is much easier to miss a weak broad component in [OII] than in [OIII], giving the impression that the [OII] is narrower or lacks kinematic structures present in [OIII]. To remedy this problem, in what follows we use only the emission lines detected with peak S/N$>$10 (about 230 objects), and we prefer S/N$>$20 (about 140 objects).

We also need to take into account the doublet nature of [OII]$\lambda\lambda$3726,3729. Because the velocity spacing of the doublet (220 km/sec) is smaller than the typical line widths, we usually cannot deblend the two components. Thus observationally the most robust course of action is to measure the non-parametric width of the entire doublet which is then expected to be slightly higher than the width of a single (non-doublet) line with the same kinematic structure. To estimate the magnitude of this bias, we take the best multi-Gaussian fits for [OIII]$\lambda$5007 for our entire sample and we simulate noiseless [OII]$\lambda\lambda$3726,3729 of the exact same kinematic structure assuming a 1:1, 1:1.2 and 1:1.4 line ratios within the doublet. We then refit the resulting [OII] profile, calculate its non-parametric widths without deblending and compare it with the `true' input [OIII] non-parametric widths. We find the following relationships: 
\begin{eqnarray}
w_{50, {\rm apparent}}{\rm [OII]}=\sqrt{w_{50,{\rm true}}^2+(175{\rm km/sec})^2};\nonumber\\
w_{80, {\rm apparent}}{\rm [OII]}=\sqrt{w_{80,{\rm true}}^2+(290{\rm km/sec})^2};\nonumber\\
w_{90, {\rm apparent}}{\rm [OII]}=\sqrt{w_{90,{\rm true}}^2+(340{\rm km/sec})^2}. \label{eq:woii}
\end{eqnarray}
The accuracy of these fitting formulae is 7\% (standard deviation). Now when we have the observed [OII] doublet, we need only measure its overall non-parametric measures and then invert equations (\ref{eq:woii}) to correct for the doublet nature of the [OII] line. The magnitude of relative asymmetries and the kurtosis parameter $r_{9050}$ of the simulated doublets are both slightly smaller than those of the input [OIII]$\lambda$5007 profiles, as expected. 

\begin{figure}
\centering
\includegraphics[scale=0.45, clip=true, trim=0cm 10cm 0cm 0cm]{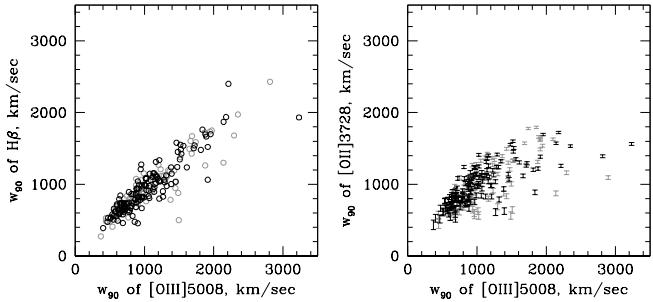}
\caption{Comparison between non-parametric width measurements for [OIII]$\lambda$5007, H$\beta$ and [OII]$\lambda\lambda$3726,3729. In light grey are sources with estimated peak signal-to-noise of H$\beta$ (left) and [OII] (right) between 10 and 20; in black are sources with S/N$>20$. Because of the minimal S/N requirement, 223 sources appear in the left panel and 237 in the right. For [OII], the top of each bar corresponds to the non-parametric measure of the width of the entire non-deblended doublet, whereas the bottom of the bar includes the correction for the doublet splitting according to eq. (\ref{eq:woii}).}
\label{pic_width}
\end{figure}

\begin{figure*}
\centering
\includegraphics[scale=0.9, clip=true, trim=0cm 10cm 0cm 0cm]{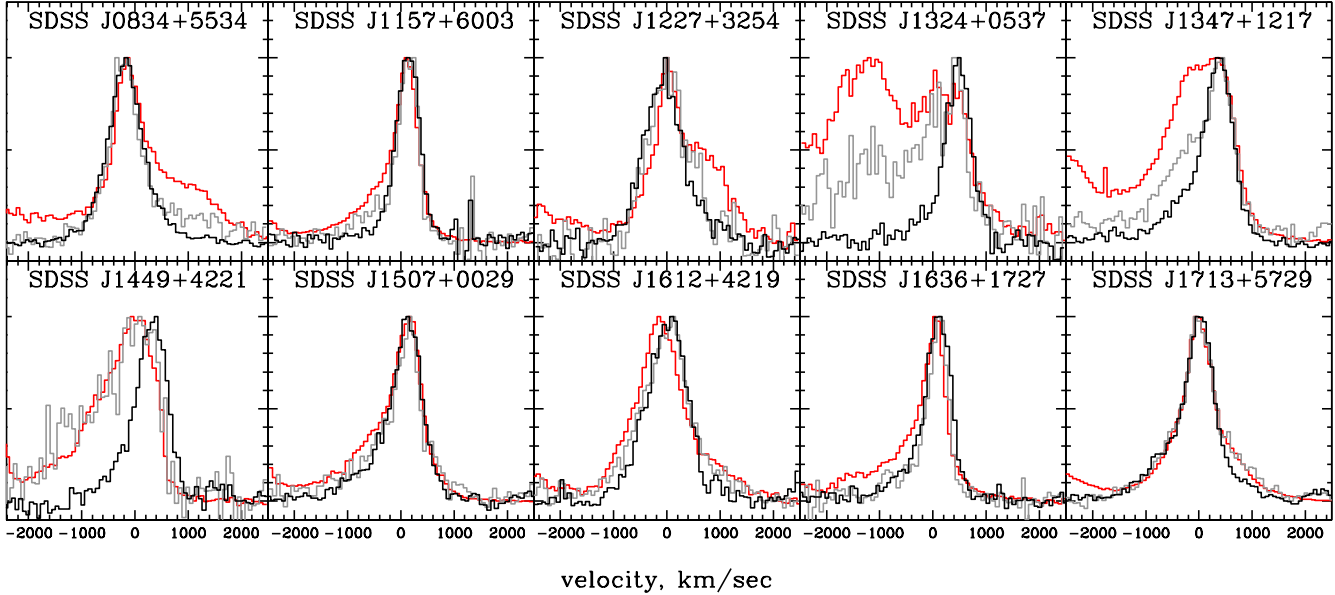}
\caption{[OIII] (red in online version and dotted grey in printed version), H$\beta$ (solid grey) and [OII] (black) line profiles for the ten objects with the broadest [OIII] which also have S/N([OII])$>$10. The [OII] profiles are as observed, without any accounting for the doublet nature of the profile; at these extreme line width the doublet splitting $\sim 220$ km/sec is not noticeable. Both H$\beta$ and [OII] are capable of showing asymmetries and complicated profile features. The kinematics of H$\beta$ tends to be less extreme than that of [OIII], and of [OII] even less so, with H$\beta$ typically seen in between [OIII] and [OII] profiles.}
\label{pic_wexamples}
\end{figure*}

In Figure \ref{pic_width}, we show the comparison between [OIII], [OII] and H$\beta$ line widths. H$\beta$ is slightly (8\%) systematically narrower than [OIII] on average. Much of the difference is likely attributable to the signal-to-noise effect described in Section \ref{sec:robust}, since it is easier to miss a weak broad component in a noisy profile of H$\beta$ than in the much higher S/N profile of [OIII]. There are only a few cases where the ``by eye'' examination of the [OIII] and H$\beta$ profiles superposed on one another reveals that the [OIII] is genuinely significantly broader than the noisier H$\beta$ (Figure \ref{pic_wexamples}). 

In contrast, [OII] is noticeably narrower than [OIII] for high-width objects (Figure \ref{pic_width}). The average doublet-corrected width of [OII] over all 237 objects with peak S/N[OII]$>10$ is only 5\% smaller than that of [OIII]; however, when we consider only objects with $w_{90}$[OIII]$>1500$ km/sec, the difference in width increases to 27\%. Overall while [OII] sometimes displays asymmetries and complicated profiles, it does not show the extremely broad features seen in [OIII], with H$\beta$ demonstrating kinematics that are intermediate between [OII] and [OIII]. 

We conduct a similar examination of He II $\lambda$4686 and [OIII]$\lambda$4363 profiles, but only a handful of objects with S/N$>10$ in these lines have relatively broad [OIII]. In these objects, [OIII]$\lambda$5007, He II $\lambda$4686 and [OIII]$\lambda$4363 kinematic structures look, within the uncertainties, consistent with one another. 

Line flux ratios involving [OIII]$\lambda$5007, H$\beta$, [OII]$\lambda\lambda$3726,3729, He II $\lambda$4686 and [OIII]$\lambda$4363 \citep{oste06, liu13a} are diagnostic of the ionization state and temperature of the gas. We look for trends in these flux ratios as a function of all kinematic parameters and luminosity of [OIII]. Unfortunately, the variations in these line fluxes are subtle enough that high S/N values in the fainter lines are required to measure their fluxes to the required accuracy. In particular, if it is the weak broad components that vary as a function of kinematics and / or luminosity, a S/N ratio of several tens in these lines would be required to detect these trends, as shown in Figure \ref{pic_noise}. As a result, we do not find any definitive trends in any of these ratios even when restricting the analysis to the small number of the highest signal-to-noise objects. Instead, we perform such measurements in Section \ref{sec:compo} using composite spectra. 

\section{Kinematics and multi-wavelength properties}
\label{sec:multiwv}

\subsection{Kinematic indicators and radio emission}
\label{sec:radio}

We cross-correlate our sample within 2\arcsec\ against the Faint Images of the Radio Sky at Twenty Centimeters (FIRST) survey at 1.4 GHz \citep{beck95}, which has typical $5\sigma$ sensitivity of $\sim 1$ mJy. When FIRST coverage is not available (about 6\% of objects), we use the NRAO VLA Sky Survey (NVSS) survey at the same frequency \citep{cond98}, which has typical $5\sigma$ sensitivity of $\sim 2.5$ mJy. Out of the 568 objects in our sample, 386 have radio detections above the catalog sensitivity. For every source, we calculate the k-corrected radio luminosity at rest-frame 1.4 GHz,
\begin{equation}
\nu L_{\nu}=4 \pi D_L^2 \nu F_{\nu}(1+z)^{-1-\alpha},
\end{equation}
where $\nu=1.4$ GHz, $F_{\nu}$ is the observed FIRST / NVSS flux (which corresponds to an intrinsically higher frequency in the rest-frame of the source), and $\alpha$ is the slope of the radio spectrum ($F_{\nu}\propto \nu^{\alpha}$), assumed to be $-0.7$. 

The exact shape of the radio luminosity function of active nuclei remains a topic of much debate, including whether there is a very broad distribution of intrinsic luminosities or whether there is a true dichotomy in this property \citep{kell89, xu99, ivez02, jian07a, cond13}. In any case, the objects on the high end are traditionally called ``radio-loud'' and are sometimes very extended in the radio band. In these cases, collimated relativistic jets propagate out to several hundred kpc from the host galaxy, and the radio emission of these sources is dominated by the lobes where the jet energy finally dissipates. The majority of our matches are weak (a few mJy) point sources at the 5\arcsec\ resolution of the FIRST survey. Since the optical broad-band emission of type 2 quasars is a poor proxy for their luminosity, we use the distribution of our sources in the [OIII] luminosity / radio luminosity plane to define the radio-loud / radio-quiet boundary \citep{xu99} and find that about 10\% qualify as classical radio-loud sources, and a similar fraction are significantly ($\ga 10$ kpc) extended \citep{zaka04}. For Seyfert galaxies, \citet{hopeng01} find that some radio-loud candidates can be missed when using the spatially integrated optical or infrared luminosity in defining the radio-loud / radio-quiet boundary and suggest using the nuclear properties in such definitions. Luckily in our case the quasars are extremely luminous and dominate the bolometric output, so our definition is unlikely to be affected by this bias. 

\begin{figure*}
\centering
\includegraphics[scale=0.9, clip=true, trim=0cm 10cm 0cm 0cm]{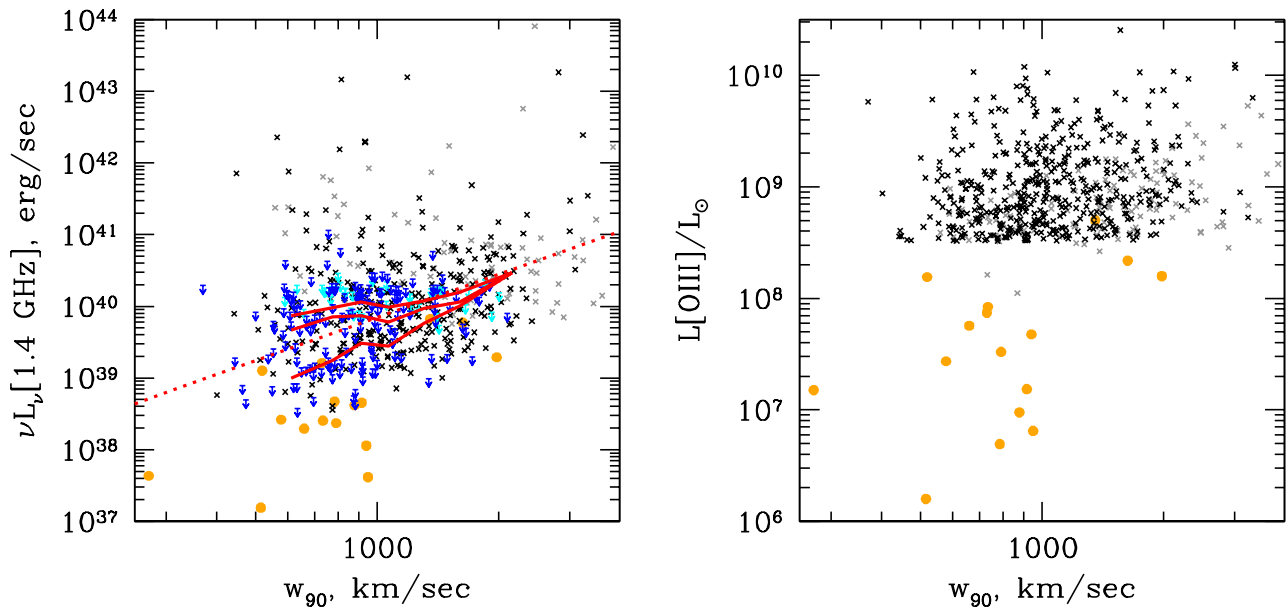}
\caption{Left: Radio luminosities (k-corrected to the rest-frame 1.4 GHz) vs the velocity width $w_{90}$ of the [OIII] emission. Crosses and upper limits show 568 type 2 quasars in our kinematic analysis. The 568 type 2 quasars in our kinematic analysis are color-coded by robustness of line fits (light grey, light blue for peak S/N([OIII])$<20$, black, dark blue for higher S/N), with crosses for radio detections and arrows for radio upper limits. Orange circles show Seyfert galaxies from \citet{veil91b}; NVSS is used to determine the radio fluxes of these sources. The thick solid red lines show median radio luminosities in bins of line width. The top line assumes that radio non-detections are close to the limit of the survey, the bottom line assumes that non-detections are ten times fainter than the limit of the survey, and the middle line assumes that the non-detections are two times fainter than the limit of the survey. The dotted line shows $\nu L_{\nu}$[1.4GHz]$\propto w_{90}^2$; it is not a fit to the data. Right: There is a slight tendency for objects with broader [OIII] to also have higher [OIII] luminosity, but because of the cutoff of the [OIII] luminosity the significance of this relationship is difficult to establish.}
\label{pic_radio}
\end{figure*}

We find a strong correlation between the [OIII] line width and the radio luminosity (Figure \ref{pic_radio}; Table \ref{tab:pnh}). For nearby lower-luminosity active galaxies, similar relationships were previously reported by many authors \citep{heck81, wils85, whit92b, nels96} and by \citet{veil91b} whose sample is shown in Figure \ref{pic_radio} for comparison. While there is a small tail of objects with very high radio luminosities, most of the correlation is due to the cloud of points at $\nu L_{\nu}[1.4{\rm GHz}]=10^{39}-10^{41}$ erg/sec. Although these radio luminosities seem high by comparison to those of the local active galaxies (e.g., the red points from the \citealt{veil91b} sample), we need to keep in mind that the [OIII] luminosities of our type 2 quasars are at the extreme end of the luminosity distribution (right panel of Figure \ref{pic_radio}). Thus in the space of [OIII] vs radio luminosities \citep{xu99} most of the type 2 quasars in our sample follow the radio-quiet, rather than the radio-loud, locus. Recently, \citet{mull13} also found a trend of increasing line width with radio luminosity using composite spectra of type 1 quasars with a wide range of [OIII] luminosities which overlaps with ours. Similarly, \citet{spoo09} reported a correlation between outflow signatures in mid-infrared emission lines and radio luminosities among ultraluminous infrared galaxies. Because these objects tend to be more dust obscured on galaxy-wide scales and have high rates of star formation which contributes to their radio emission, a direct comparison between our sample and theirs is complicated, but at face value the radio luminosities of the two samples do overlap. 

Taking all sources and radio upper limits at face value (i.e., assuming that the non-detections are close to the survey limit) yields a Spearman rank correlation coefficient of $r_{\rm S}=0.29$. Excluding radio-loud sources using their distribution in the [OIII]-radio plane \citep{zaka04} yields $r_{\rm S}=0.33$. If the non-detections are significantly below the survey limit, then the correlation is even stronger because there are hardly any non-detections on the high-$w_{90}$ end of the diagram. In Section \ref{sec:origin} we demonstrate that the radio fluxes of objects not detected by FIRST are likely within a factor of two of the FIRST survey limit. If we suppress all upper limits by a factor of two (but keep radio-loud sources in), the correlation has $r_{\rm S}=0.34$. 

We also find an anti-correlation between radio luminosity and absolute asymmetry, $r_{\rm S}=-0.37$, in the sense that objects with stronger blue asymmetry tend to have stronger radio emission. A similar relationship exists between radio luminosity and the median velocity $v_{50}$. In all these cases the null hypothesis (that the two datasets are uncorrelated) is rejected with $P_{\rm NH}<10^{-5}$. Radio luminosity is not correlated with the other kinematic measures of outflow activity (relative asymmetry and kurtosis $r_{9050}$). For comparison, we also show the weaker trend between $w_{90}$ and [OIII] luminosity reported in Section \ref{sec:analysis} in the right panel of Figure \ref{pic_radio}. 

\subsection{Kinematic indicators and infrared luminosity}

We cross-correlate the entire sample of SDSS type 2 quasars \citep{reye08} against the Wide-field Infrared Survey Explorer (WISE) catalog within 6\arcsec. Out of 887 objects in the catalog, 876 objects have matches in W1 (3.6\micron) and W2 (4.5\micron); 829 objects in W3 (12\micron); and 773 objects in W4 (22\micron) with signal-to-noise ratio above 2.5. The 11 objects without W1 matches are visually examined; in almost all cases there is an actual detection at the position of the quasar, but it is blended with a brighter nearby object and is thus not reported in the catalog. We interpolate between the WISE fluxes using piece-wise power-laws to calculate $\nu L_{\nu}$ at rest-frame 5 and 12 \micron\ and the index between these two, $\nu L_{\nu}\propto \lambda^{\beta}$ (higher index means redder spectral energy distribution). As the majority of the sources are well above the detection limit for the survey, the analysis of the WISE matches is not affected by non-detections to the same extent as the analysis of radio emission in the previous section. 

\begin{figure*}
\centering
\includegraphics[scale=0.9, clip=true, trim=1cm 11.5cm 0cm 0cm]{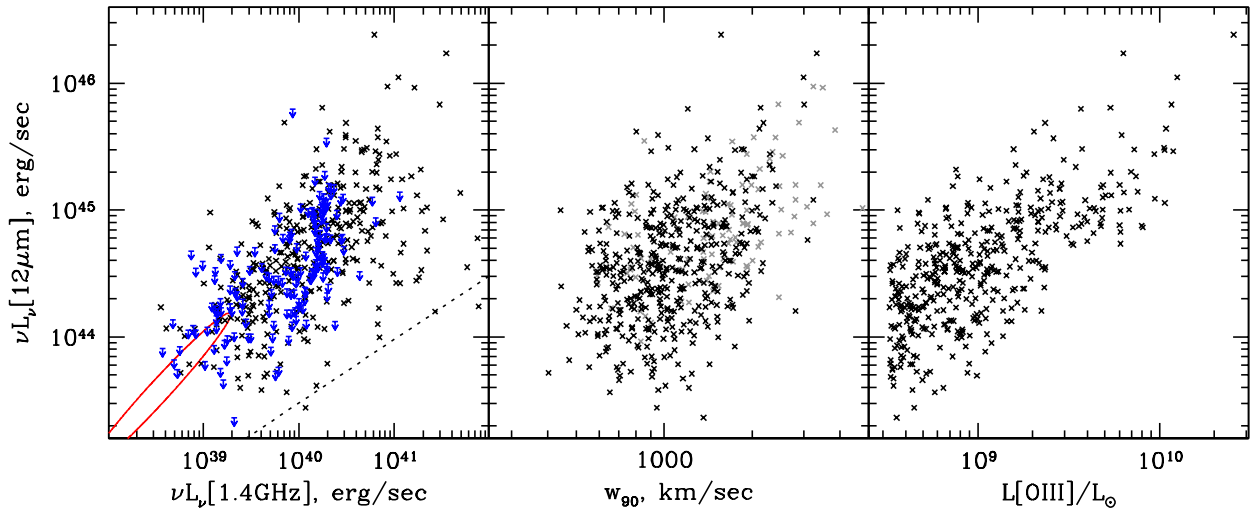}
\caption{WISE monochromatic k-corrected luminosities as a function of radio luminosities (left), [OIII] velocity widths (middle) and [OIII] luminosities (right). In the left panel, the blue points are those with upper limits on radio fluxes, whereas the black points are radio detections. The red ellipse is the locus of star-forming galaxies from \citet{rosa13}, and the dashed line is their separation line between infrared-bright (above the line) and infrared faint (below the line) branches of active nuclei. In the middle panel, grey points are for objects with peak S/N[OIII]$<20$.}
\label{pic_wise}
\end{figure*}

The mid-infrared luminosities of type 2 quasars in our sample strongly correlate with their radio luminosities, [OIII] velocity widths and [OIII] luminosities (Figure \ref{pic_wise}). The relationships between infrared luminosities, radio luminosities and [OIII] luminosities in low-luminosity active galaxies have been pointed out by many authors before, most recently by \citet{rosa13}. For direct comparison with their work, we show the locus of normal starforming galaxies and the \citet{rosa13} line separating two branches of active nuclei in the left panel of Figure \ref{pic_wise}. Using $\nu L_{\nu}$ at 5\micron\ instead results in similar relationships albeit with somewhat larger scatter. 

Mid-infrared luminosities $\nu L_{\nu}$[12\micron] are correlated with all kinematic measures, in the sense of higher mid-infrared luminosity in objects with stronger outflow signatures, with significance ranging between $P_{\rm NH}=0.02$ and $<10^{-5}$ (Table \ref{tab:pnh}). The strongest correlation is with $w_{90}$ ($r_{\rm s}=0.44$, $P_{\rm NH}<10^{-5}$). There is a hint ($P_{\rm NH}=0.04$) that the mid-infrared luminosity correlates positively with positive values of $v_{50}$, while correlating negatively with the negative values, suggesting that either a strong blue asymmetry or a strong red asymmetry may be a sign of an outflow. Objects with higher $w_{90}$ tend to have redder mid-infrared spectral energy distributions (higher $\beta$), with $r_{\rm S}=0.23$ and $P_{\rm NH}<10^{-5}$. 

\citet{rosa13} point out that the Seyfert galaxies in their sample lie almost exactly on top of the locus of the star-forming galaxies in the radio / infrared diagrams. These authors conclude that only 15\% of the infrared flux of these objects is due to the active nucleus and that the correlation between infrared and radio fluxes seen among Seyfert galaxies is simply a reflection of the standard radio / infrared correlation due to star formation. Type 2 quasars appear to lie on the luminous extension of the locus of the star forming galaxies, and thus it is tempting to postulate that the same arguments apply in our sample, except the star formation rates of the host galaxies must be much higher than those seen in Seyferts by \citet{rosa13}. 

However, this explanation is unlikely to extend to the objects in our sample. The mid-infrared colors and fluxes of type 2 quasars at these luminosities are dominated by the quasar, not by the host galaxy \citep{lacy04, ster05, zaka08}. Thus the strong correlation between radio and mid-infrared in this regime (and the excess of the radio emission over the amount seen in nearby star-forming galaxies) suggests that the radio emission in radio-quiet quasars is related to the quasar activity, not to the star formation in its host. 

Our sample has 54 objects in common with \citet{jia13} who analyzed XMM-Newton and Chandra snapshots of a large sample of obscured quasars deriving their X-ray luminosities, spectral slopes and amount of intervening neutral gas absorption. These objects were either targeted by X-ray observatories or serendipitously lie in the fields of view of other targets. Similarly to \citet{veil91b}, we do not find any correlations between any of the kinematic indicators and any of the X-ray spectral fitting parameters. In particular, there is no correlation between the optical line width and the absorption-corrected (intrinsic) X-ray luminosity. Removing the 24 Compton-thick candidates (in which obtaining intrinsic X-ray luminosities is particularly difficult) still reveals no relationship between X-ray parameters and gas kinematics. 

One possibility is that the lack of such correlation implies the lack of strong influence of X-ray emission on the launching of the winds. Another possibility (which we find more likely) is that the existing X-ray observations of obscured quasars are not yet of sufficient quality to probe this relationship. The uncertainties in the intrinsic X-ray luminosities are rather high, because the observed fluxes need to be corrected for intervening absorption. As a result, even the correlation between X-ray and mid-infrared luminosities -- which are both supposed to be tracers of the bolometric luminosity -- is rather weak ($P_{\rm NH}\simeq 0.02$). Among the 30 Compton-thin sources, the X-ray to mid-infrared luminosity ratio has a dispersion of 0.7 dex, similar to the value reported for local Seyfert 2 nuclei \citep{lama11} who argue that correcting for intrinsic absorption is difficult even when high-quality X-ray observations are available. 

We perform a simulation in which we randomly draw 30 points from the infrared vs kinematics correlation in Figure \ref{pic_wise}, middle. We find that the correlation is still detected with $P_{\rm NH}\la 0.01$ significance. But if we add a Gaussian random variable with an 0.7 dex dispersion to the log of the infrared luminosity, the correlation is no longer detected. Thus either an intrinsic dispersion or observational uncertainties (related to difficulties of correcting for intervening absorption) of this magnitude are sufficient to destroy a correlation in a sample of 30 objects (the number of Compton-thin obscured quasars with available X-ray luminosities). It will be interesting to probe the connections between the ionized gas kinematics and ultra-violet, optical and X-ray luminosities in type 1 quasars, where correcting for intervening absorption is not a significant problem and where the relative strengths of these correlations could elucidate the primary driving mechanism of the ionized gas outflows. 

\section{Composite spectra}
\label{sec:compo}

\subsection{Constructing composites}

\begin{figure*}
\centering
\includegraphics[angle=-90,scale=0.65]{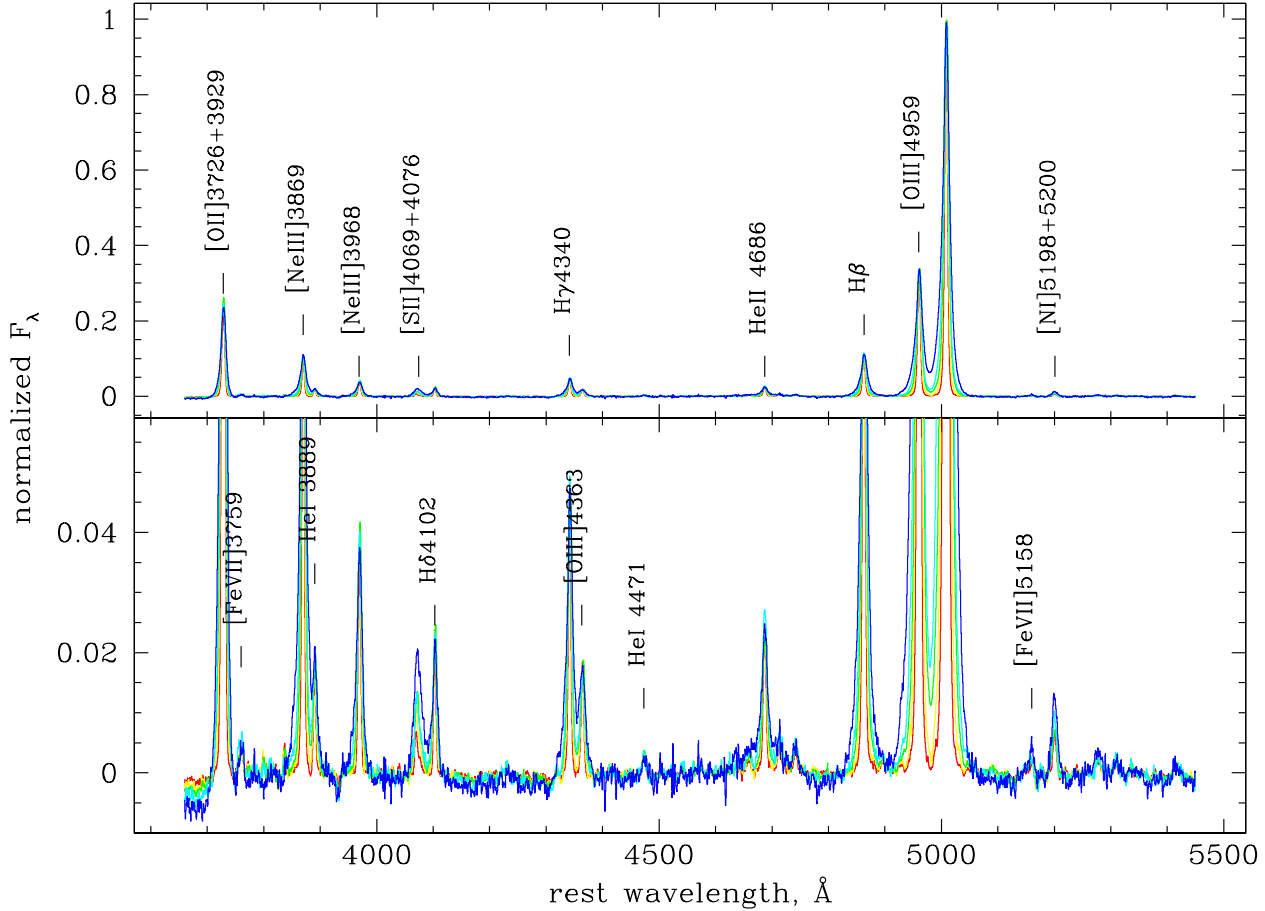}
\caption{Composite in 5 bins of [OIII] $w_{80}$ width (from red to blue in online version or light to dark in printed version $w_{80}$ is increasing) normalized to the peak flux density of the [OIII] flux. Composite is straight-up error weighted mean in the host frame (whether or not the host subtraction is reported to be accurate). The bins are $w_{80}=284-546$ km/sec, $546-673$ km/sec, $673-856$ km/sec, $856-1153$ km/sec, $1153-2918$ km/sec. Most of the line peaks line up from one composite to the next, but the peaks of [SII]$\lambda\lambda$4069,4076 and [NI]$\lambda\lambda$5198,5200 show an increase. [FeVII]$\lambda$5158 appears to increase as well.}
\label{pic_comp_width}
\end{figure*}

\begin{figure*}
\centering
\includegraphics[angle=-90,scale=0.65]{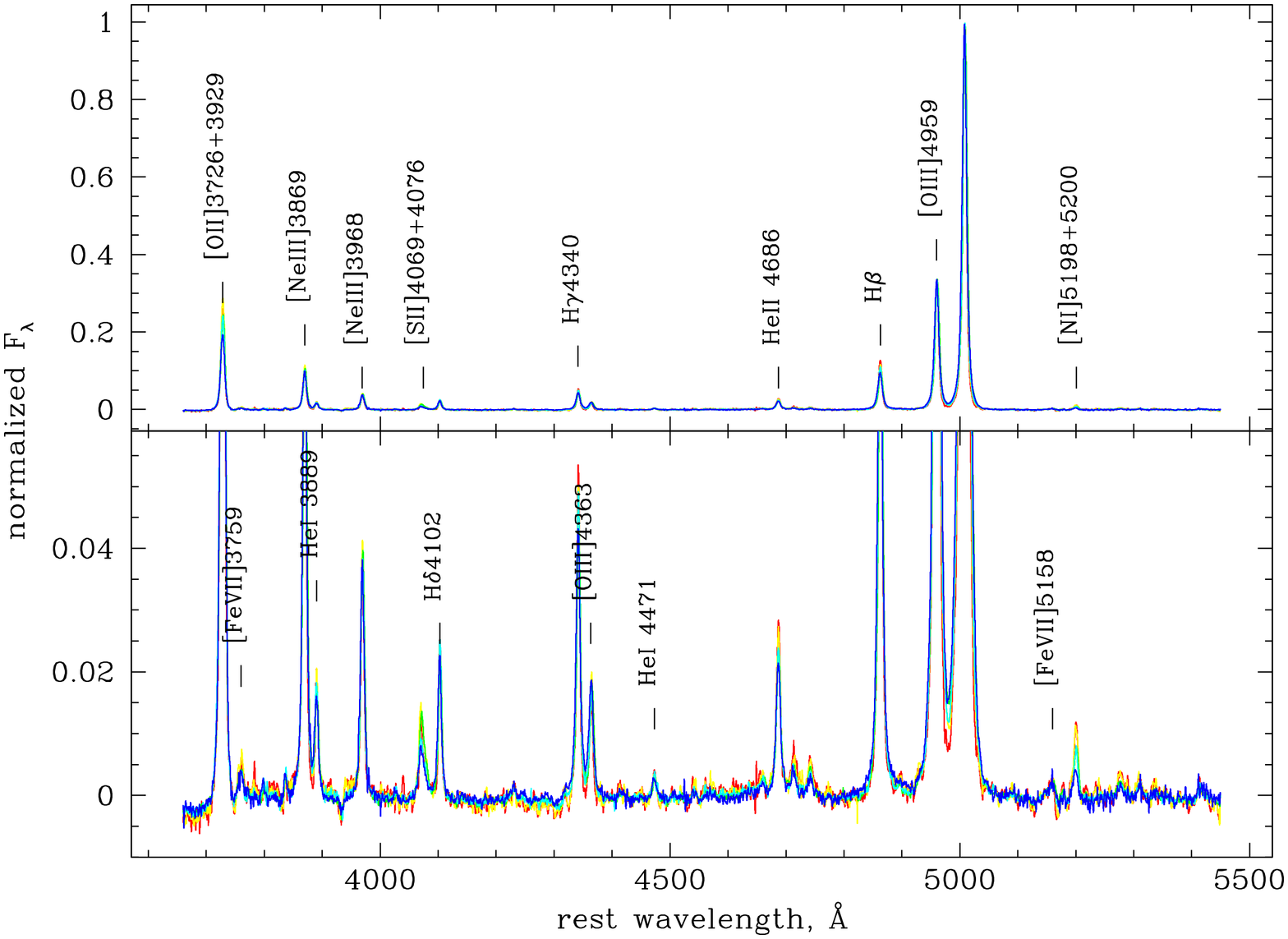}
\caption{Composite in 5 bins of [OIII] luminosity (from red to blue in online version or light to dark in printed version $L$[OIII] is increasing) normalized to the peak flux density of the [OIII] flux. The bins are $\log L$[OIII]$/L_{\odot}=8.50-8.61$, $8.61-8.78$, $8.78-8.97$, $8.97-9.23$, $9.23-10.41$. Many of the line peaks decrease in the red$\rightarrow$ blue sequence, but the line widths do not vary drastically, reflecting the weakness of the relationship between [OIII] luminosities and widths.}
\label{pic_comp_lum}
\end{figure*}

To further test the trends we find in Sections \ref{sec:optical} and \ref{sec:multiwv} and to study weak emission features, we produce several sets of composite spectra. We choose a quantity that is easily measurable in every object (e.g., [OIII] line width in our first example) and bin the sample into five equal-size bins (114 objects) in this quantity. We then arithmetically average all host-subtracted spectra within each bin. This allows us to obtain high signal-to-noise composites while being able to tease out the dependencies on the chosen parameter. The composites produced in five bins of the [OIII] velocity width are presented in Figure \ref{pic_comp_width} and the composites in five bins of [OIII] luminosity in Figure \ref{pic_comp_lum}. Below we also discuss composites made in bins of infrared and radio luminosity, although these are not shown. 

Even though the spectra are already host-subtracted, for accurate measurements of very weak lines we need a more accurate continuum subtraction. We select a dozen continuum-dominated wavelength intervals and spline-interpolate between them to produce the model of the continuum which is then subtracted from the composite spectrum. Continuum subtraction is the dominant source of systematic uncertainty in measuring the weakest lines. 

We then construct the velocity profile of the narrow-line region from the [OIII]$\lambda\lambda$4959,5007 lines and we use this profile to fit about 15 emission lines. Several doublets which have wavelength separations that are too small to be resolved in our spectra are fitted with fixed ratios between the components, as follows: [OII]$\lambda\lambda$3726,3729 with a 1:1 ratio, [SII]$\lambda\lambda$4069,4076 with a 3:1 ratio (the doublet structure of this line is clearly visible in the lower $w_{80}$ composites), and [NI]$\lambda\lambda$5198,5200 with a 1:1 ratio. For each emission feature, given the velocity profile from the [OIII] line, there is only one adjustable parameter -- its amplitude. The fit is linear in all 15 amplitudes. 

A few features are close blends: He I $\lambda$3889 is blended with H$\zeta$ (8$\rightarrow$2 transition), and [NeIII]$\lambda$3968 is blended with H$\varepsilon$ (7$\rightarrow$2 transition). In both cases, the non-hydrogen emission makes a larger contribution to the blend, but not an overwhelmingly dominant one. To measure He I $\lambda$3889 and [NeIII]$\lambda$3968, we estimate the Balmer decrement from H$\beta$, H$\gamma$ and H$\delta$ and use the derived extinction values to estimate H$\varepsilon$ and H$\zeta$, assuming Case B recombination \citep{oste06}. We then subtract the extrapolated H$\varepsilon$ and H$\zeta$ fluxes from the corresponding blends to obtain He I $\lambda$3889 and [NeIII]$\lambda$3968 fluxes separately. 

The values of extinction we find using Balmer decrement and the Small Magellanic Cloud extinction curve from \citet{wein01} are in the range $A_V=1.0-1.5$ mag, in agreement with our previous estimates for type 2 quasars \citep{reye08} and with typical values in the literature for narrow-line regions of Seyfert galaxies \citep{benn06}. Extinction values are higher ($A_V\simeq 1.5$ mag) for the two highest width composites than for the other ones ($A_V\simeq 1.0-1.1$ mag), somewhat reminiscent of the results by \citet{veil91a} who finds higher extinction values for objects with stronger line asymmetries. As a function of luminosity, extinction decreases steeply and monotonically, from $A_V=1.7$ mag to 0.9 mag. Because it is not clear that the extinction values derived from Balmer decrements apply to all other emission features (which may originate in a different spatial region), we do not apply extinction correction to any measurements, unless explicitly stated otherwise. 

The optimal composite-making practices depend on the goal of the composite \citep{vand01, lacy13}. The typical per-spectroscopic-pixel errors of the SDSS spectra are the result of the overall plate reductions and thus are not too dissimilar from one object to the next, so in an error-weighted average all weights would be roughly the same. Therefore we chose to use simple arithmetic averages to produce our composites \citep{lacy13}. It is encouraging that the [OIII]/[OII] line ratios measured from the composites ($\log$[OIII]/[OII]$=0.67-0.74$, depending on [OIII] width) are consistent with the one measured from individual spectra (mean=median and sample standard deviation $0.70\pm 0.23$) and [OIII]$\lambda$4959/[OIII]$\lambda$5007 is within 1\% of its theoretical value in all composites. This means that our composite-making procedure does a reasonable job of preserving line ratios, both for cases when the intrinsic distribution of the line ratio is very narrow (e.g., [OIII]$\lambda$4959/[OIII]$\lambda$5007) and when it is fairly broad (e.g., [OIII]/[OII]).

\subsection{Wolf-Rayet features}

One striking result is the appearance of a broad complex around He II $\lambda$4686 in the composite with the highest $w_{80}$. This feature is somewhat reminiscent of the broad emission signatures of Wolf-Rayet stars \citep{brin08, liu09} -- massive young stars which can radiatively drive outflows from their photospheres with velocities reaching 2000 km/sec. A detection (or lack thereof) of young stellar populations is of particular importance in our work, since supernova explosions are capable of driving powerful outflows. Thus, a significant star-forming population could be responsible for at least some of the outflow activity, and therefore this feature deserves particular scrutiny.  

The emission-line region photo-ionized by the quasar produces a multitude of forbidden and recombination (`nebular') features in the wavelength range between 4600\AA\ and 4750\AA\ \citep{liu09}. The broad lines produced by the Wolf-Rayet stars at these wavelengths are N V $\lambda$4613, N III $\lambda$4640, C III/IV $\lambda$4650, and He II $\lambda$4686 \citep{brin08}. Thus the challenge is to disentangle the two sets, made even more difficult by the fact that at the high velocities of the ``narrow'' lines in our highest $w_{80}$ objects the velocities in the quasar-ionized set are not dissimilar from the velocities in the Wolf-Rayet set. 

\begin{figure*}
\centering
\includegraphics[scale=0.85, clip=true, trim=0cm 14.5cm 0cm 0cm]{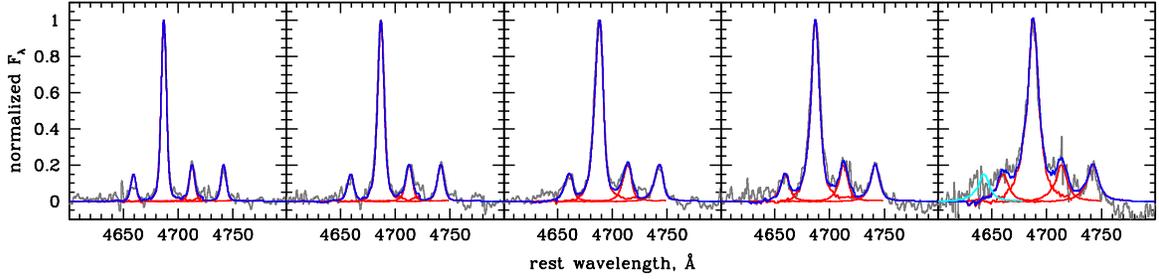}
\caption{Zoom in on the wavelength region that may contain Wolf-Rayet signatures for the five composite spectra made in bins of [OIII] line width, which is increasing from left to right. In each panel, the grey histogram shows the composite spectrum and the red curves (light grey in printed version) show H$\beta$ velocity profiles from the same composite scaled to match the amplitudes of [Fe III]$\lambda$4658, He II $\lambda$4686, [Ar IV]$\lambda$4711 and [Ar IV]$\lambda$4740. In all five panels, [Fe III]/He II=0.15 and each [Ar IV]/He II=0.2. The blue curve (solid black in the printed version) shows the sum of the red profiles. The light blue curve (dotted in the printed version) in the last panel shows the putative N III $\lambda$4640 component.}
\label{pic_wr}
\end{figure*}

In Figure \ref{pic_wr} we zoom in on the wavelength range of interest. We find that only four nebular emission features ([Fe III]$\lambda$4658, He II $\lambda$4686, [Ar IV]$\lambda$4711 and [Ar IV]$\lambda$4740) are sufficient to explain most of the observed flux. For each composite, we assume that these four features are associated with the narrow-line region of the quasar and have the same kinematic structure as H$\beta$. We take the H$\beta$ velocity profile from the same composite and we adjust its amplitude to fit each of these four features. We see no evidence that the relative fluxes of the four lines vary from one composite to the next; thus the models shown in Figure \ref{pic_wr} have [Fe III]/He II fixed at 0.15 and each [Ar IV]/He II fixed at 0.20, making it essentially a one-parameter fit (the overall ratio of He II to H$\beta$). No broad component in He II is required to produce an adequate fit. There is some evidence in the last composite of some emission filling in between the two argon features, possibly due to [Ne IV]$\lambda$4725. 

Although most of the ``broad'' feature turns out to be due to a superposition of relatively broad nebular lines, in the last panel we see excess emission centered at around 4640\AA. This is close to the wavelength of one of the more prominent Wolf-Rayet features, N III $\lambda$4640. To estimate the flux of this feature, we model it using the same H$\beta$ velocity profile (although in this case there is no particular physical reason to do so, since the N III line is expected to reflect the kinematics of the stellar winds, rather than be produced in the extended low-density medium of the host galaxy and reflect the kinematics of quasar-ionized gas). We find a reasonable fit with N III/He II=0.15, but we rather doubt the identification of this feature as N III associated with the Wolf-Rayet stars in the host galaxy. The feature is noisy and is somewhat offset to the blue from the nominal expected centroid. When all spectra are coadded together into a single composite, the evidence for N III disappears. 

We assume for the moment that the N III feature is in fact detected and estimate its median luminosity. To this end, we use the observed N III / [OIII] ratios in the composite and the median [OIII] luminosity of the objects in the composite $\log L$[OIII]/$L_{\odot}=8.91$ to find $L$(N III)$\simeq 3\times 10^6L_{\odot}$. We then use N III fluxes and starburst models from \citet{scha99} and \citet{leit99} to estimate star-formation rates of $\sim 6 M_{\odot}$/year. This is significantly lower than the estimates of star formation rates in type 2 quasar hosts by \citet{zaka08} (a few tens $M_{\odot}$/year). The difference between the two methods is not alarming because there are still significant discrepancies between modeled and observed fluxes of Wolf-Rayet features, especially as a function of metallicity \citep{brin08}. But the low fluxes of Wolf-Rayet features in our composites -- if they are detected at all -- further reinforce our understanding that the star formation rates in type 2 quasar hosts are not adequate for producing the observed radio emission. At 10 $M_{\odot}$/year, using calibrations by \citet{bell03} and \citet{rosa13} we find that typical star-forming galaxies would produce $\nu L_{\nu}$[1.4GHz]$=2.5\times 10^{38}$ erg/sec and $\nu L_{\nu}$[12\micron]$=2.9\times 10^{43}$ erg/sec, more than an order of magnitude below the values seen in our sample.  

\subsection{Line ratios as a function of line width and line luminosity}

\begin{figure}
\includegraphics[scale=0.45, clip=true, trim=0cm 0.5cm 9.5cm 0cm]{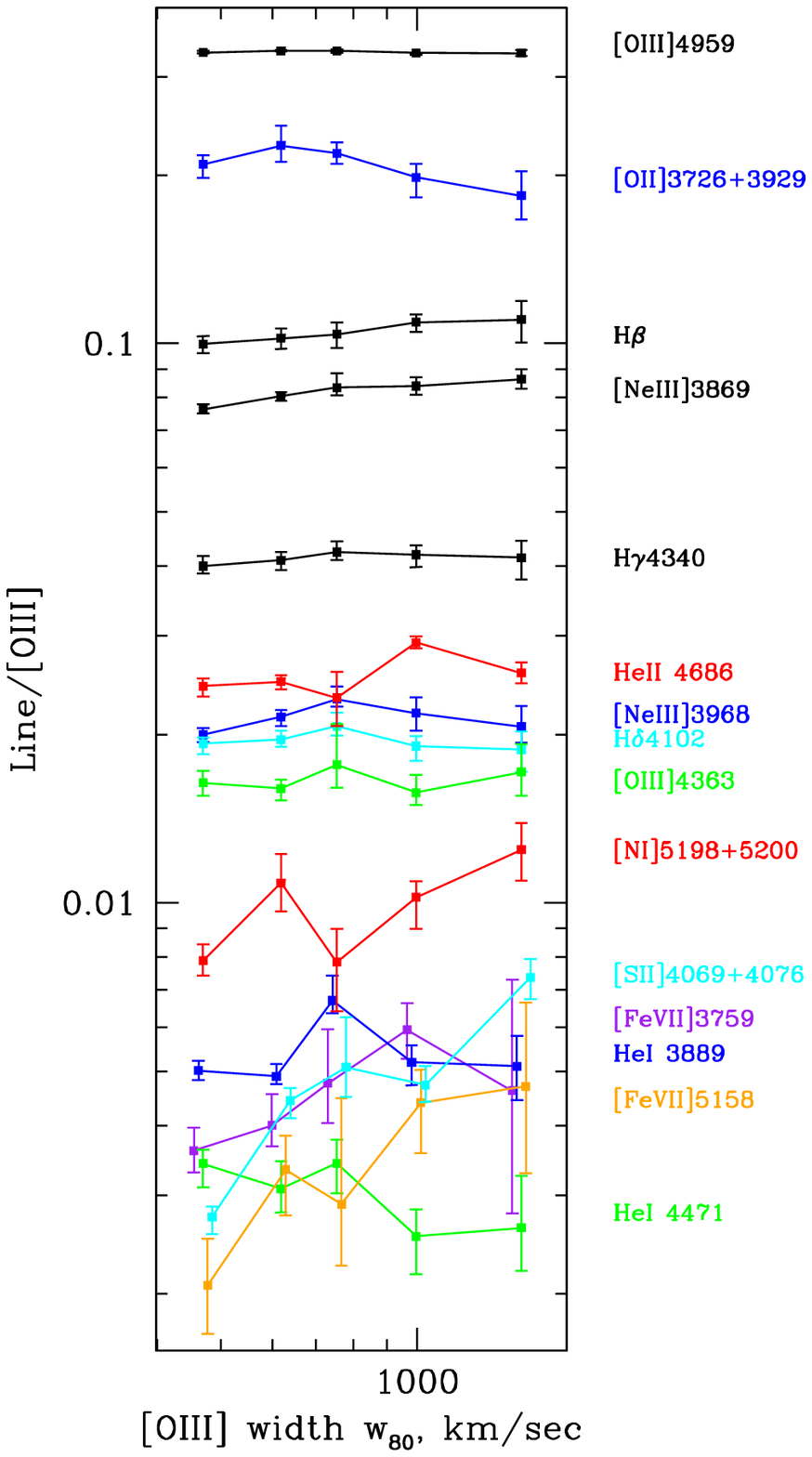}%
\includegraphics[scale=0.45, clip=true, trim=2cm 0.5cm 9.5cm 0cm]{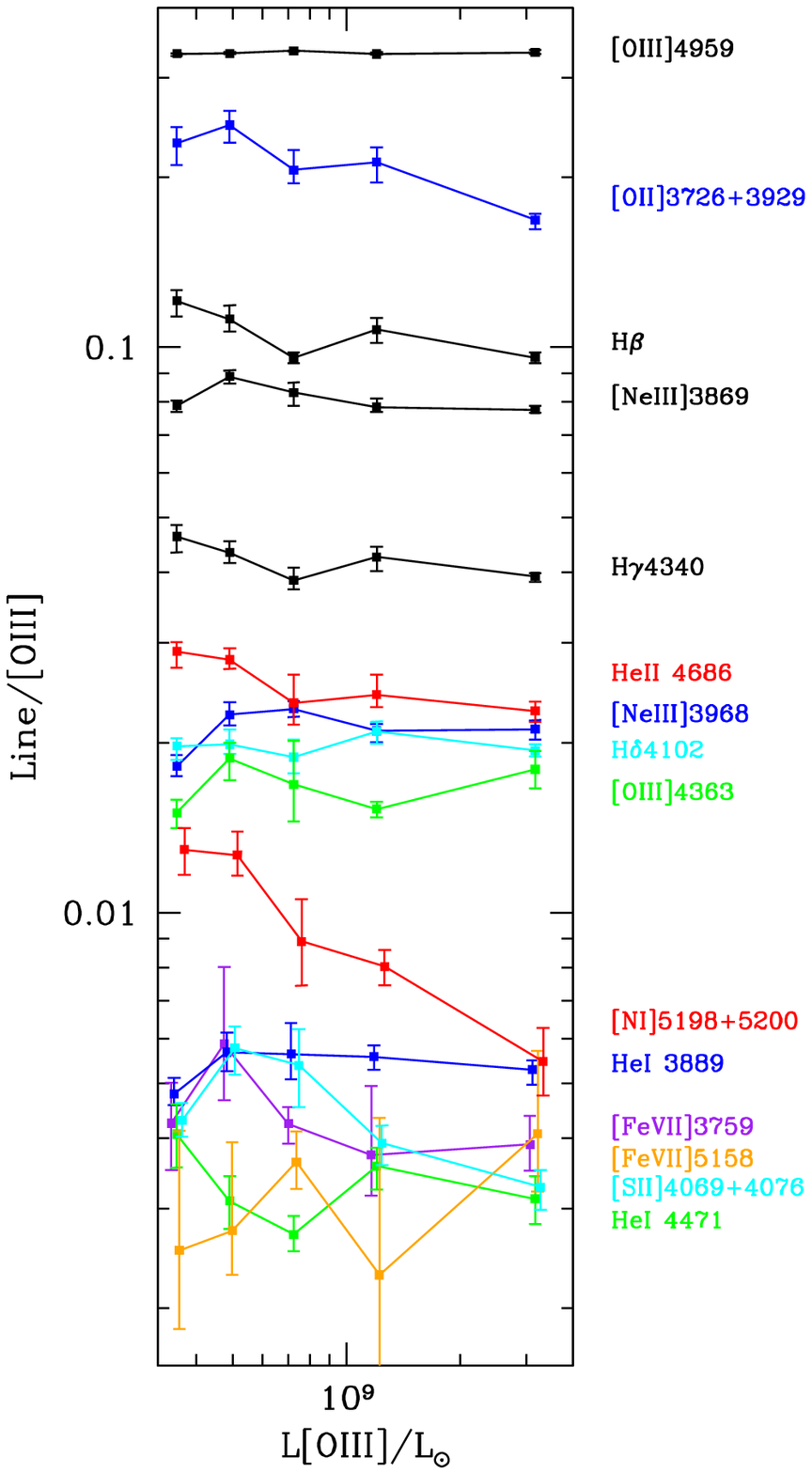}\\
\caption{Line ratios as a function of line width (left) and [OIII] line luminosity (right) calculated from the composite spectra. On the horizontal axis, we use the median $w_{80}$ (in km/sec) or the [OIII] line luminosity of the spectra contributing to the particular composite. Some curves are color-coded (or shown with different line types in the printed version) in no particular order to help distinguish them on the plot, and the bottom six curves are offset from each other by a small amount in the horizontal direction for the same reason; no actual change of line width or line luminosity is present.}
\label{pic_comp_ratio}
\end{figure}

Figure \ref{pic_comp_ratio} summarizes our analysis of the line ratios as a function of [OIII] width and luminosity. We report the trends we see among the composite spectra at face value, but we caution that many of the lines we discuss are too weak to be detected in individual objects. As a result, in most cases it is not possible at the moment to evaluate whether the trends we see in the composite spectra accurately represent the trends in the population; a couple of exceptions are noted below. 

For every composite, we compute errors in the line ratios using bootstrapping: we randomly resample (100 times) with replacement the set of spectra that contribute to each composite, recompute the composite, and recompute the line ratios. The error bars shown in Figure \ref{pic_comp_ratio} encompass 68\% of the resampled line ratios and represent the statistical error; the systematic uncertainties (those due to continuum subtraction, due to fitting procedure and due to assuming the same kinematic shape for all emission lines) are not included. Typical error bars for the line ratios of the brighter lines such as [OII] and [NeIII] are about $\sim 0.03$ dex, reaching 0.2 dex for fainter lines such as [FeVII]. For brighter lines the dispersion of line ratios within the population can be measured from individual objects and is typically $\sim 0.3-0.4$ dex \citep{ster12}; thus our bootstrapping procedure confirms that the error in the mean is consistent with $\sim 1/\sqrt{N}$ of the dispersion within the population ($N=114$ is the number of the spectra in each composite). For fainter lines measurement uncertainties dominate over the statistical ones. 

We find an increase of [SII]$\lambda\lambda$4069,4076 (and to a lesser extent, [NI]$\lambda\lambda$5198,5200) as a function of [OIII] width, which is our proxy for outflow velocity. Over the range of [OIII] widths that we probe, these lines increase respectively by factors of 2.6 and 1.6 (or by 0.4 and 0.2 dex) relative to [OIII]. Line ratios that vary with kinematics are often a tell-tale sign of shock excitation in star-burst galaxies and LINERs \citep{veil94, veil95}. [SII] and [NI] trace warm weakly ionized gas phase, which is common in shocks that penetrate deep into clouds and that can produce extended low-ionization regions \citep{tiel05}. The increase in the relative prominence of these lines suggests an increase in shock-ionization contribution, perhaps in direct response to an increase of the quasar wind velocity. 

The [OIII]$\lambda$4363/[OIII]$\lambda$5007 ratio is $\simeq 1.7\times 10^{-2}$ and does not vary in a regular fashion with either [OIII] width or luminosity. This value is significantly higher than that predicted by photo-ionization models \citep{vill08}, and is higher still if corrected for extinction, suggesting some contribution from shocks. Combined shock- and photo-ionization models \citep{moy02} can help explain this ratio, but because photo-ionization makes the dominant contribution to the [OIII] emission the lack of dependence on line kinematics is not too surprising. 

Both [SII]$\lambda\lambda$4069,4076 and especially [NI]$\lambda\lambda$5198,5200 appear to decline as a function of [OIII] luminosity, by factors 1.3 and 2.4 (0.1 and 0.4 dex), respectively. A similar decline of stronger low-ionization lines such as [NII]$\lambda$6583, [SII]$\lambda\lambda$6716,6731 and [OI]$\lambda$6300 relative to [OIII] is seen in type 1 quasars by \citet{ster12}. Unfortunately, these particular lines are not accessible in our higher-redshift objects, but applying their scalings to our lines and the range of [OIII] luminosities probed by our data, one would predict a decline by factors of 1.3$-$1.8, consistent with our observations. This decline suggests that the low-ionization regions are being destroyed as $L$[OIII] increases, likely as a result of the increase in the bolometric luminosity and thus the availability of ionizing photons. The same effect could also be produced by the increase in the opening angle of quasar obscuration, which would increase the size of the photo-ionized regions and lead to obliteration of the low-ionization regions by direct quasar radiation, but the strong positive [OIII]-infrared correlation suggests that the bolometric luminosity increase is the more likely driver for the observed correlations. 

The [OII]/[OIII] ratio decreases by a factor of 1.4 (0.14 dex) as a function of [OIII] luminosity, which was previously reported by \citet{kim06}, with our values being in agreement with theirs. Quasar photo-ionization models produce poor fits to this ratio \citep{vill08}. Explanations usually involve a combination of high-ionization regions photo-ionized by the quasar and low-ionization regions photo-ionized by star formation \citep{kim06}, so the decrease in [OII]/[OIII] with [OIII] luminosity then indicates the decrease in the relative contribution of the lower-ionization regions, which is similar to the behavior of [SII] and [NI] lines. However, unlike [SII] and [NI], [OII]/[OIII] decreases with [OIII] line width. We speculate that this trend may be an indirect hint that the quasar-driven wind of increasing velocity (as measured by [OIII] width) suppresses star formation in the host galaxy, a phenomenon which on occasion can be observed in action in individual sources \citep{cano12}. 

We do not find any noticeable trends in line ratios with radio luminosity, except that the lines are getting broader as the radio luminosity increases. This confirms our previous finding of the correlation between [OIII] width and radio luminosity (Figure \ref{pic_radio}), but suggests that radio emission is not one of the principal drivers of the state of the ionized gas. Furthermore, there is no evidence that Wolf-Rayet features appear in higher radio luminosity objects, which is in qualitative agreement with our previous conclusion that radio emission is not dominated by the star formation in the host. We also examine composite spectra made in bins of infrared luminosity and find their behavior similar to those made in bins of [OIII] luminosity, except that the width of the lines is a significantly stronger function of infrared than [OIII] luminosity, as we already know from Figures \ref{pic_radio} and \ref{pic_wise}. 

In the highest line width composite, the profile of [OII] is visibly narrower than that of [OIII], and indeed we find $w_{90}$[OIII]$=2132$ km/sec ($w_{80}=1498$ km/sec), whereas that of [OII] (corrected for doublet structure using eq. \ref{eq:woii}) is 1410 km/sec (1040 km/sec). This confirms the trend we see in Figure \ref{pic_width} and may mean that the [OII]/[OIII] ratios calculated from the composite spectra are overestimated (as they are computed with a fixed velocity profile of [OIII]). For any lines other than [OII], even with the high S/N of the composite spectrum it is not clear whether the kinematic structures are consistent from one line to the next. There is a hint that [NI]$\lambda\lambda$5198,5200 is narrower than [OIII] as well. We do not detect any obvious overall blueshifts of any features relative to [OIII]. 

Another curious tidbit is the increase in the He I $\lambda$3889 / He I $\lambda$4471 ratio with [OIII] width. This ratio ranges from 1.2 to 2.1, which is significantly lower than the standard Case B values \citep{oste06}. If we correct them for the extinction, we find values in the range 1.8 to 2.6 (still somewhat lower than Case B) and increasing monotonically with $w_{80}$ (in the last $w_{80}$ bin, He I $\lambda$3889/He I $\lambda$4471 is likely underestimated becase He I $\lambda$3889 starts to blend with [NeIII]$\lambda$3869), without any trend with $L$[OIII]. It is possible that He I is affected by self-absorption \citep{monr13} which declines as a function of line width.

\subsection{Coronal lines}

We detect two very high-ionization features, [Fe VII]$\lambda$3759 and [Fe VII]$\lambda$5158, which show a tendency to increase with [OIII] width. Such lines, termed ``coronal'', arise from ions with high ionization potential, 99 eV in the case of FeVII. The most frequently discussed mechanism for their production is excitation by quasar photo-ionization fairly close to the nucleus, akin to an ``intermediate'' region between the very compact broad-line region and a more extended narrow-line one \citep{gran78, gelb09, mull09, mull11}. In photo-ionization models, high electron and photon densities are typically required to produce coronal lines, which explains why they arise close to the nucleus, e.g., at sub-pc scale in the case of a Seyfert galaxy modeled by \citet{mull09}. These authors propose that the inner edge of the obscuring material is the plausible location of the coronal emission regions. Similarly, the rapid variability of coronal line emission in another Seyfert galaxy suggests that the size of the emission region is just a few light years \citep{komo08}. 

Such small distances present a challenge to the study by \citet{rodr11} who find that the coronal lines are detected at similar rates in type 1 and type 2 objects, indicating that they originate outside the obscuring material, although \citet{rodr11} still require very high densities, $n_e=10^8-10^9$ cm$^{-3}$. If coronal lines originate somewhere near the inner edge of the obscuring material, depending on the orientation we may see them in an occasional type 2 object, but overall one would expect to see lower detection rates of coronal lines in type 2 objects than in type 1s in this scenario. 

It would be surprising to see a region close to or inside of the obscuring material in type 2 quasars, and even more surprising to see its flux correlate with the kinematics of much lower-ionization, much more extended gas. Some recent models demonstrate the possibility of producing coronal lines on the surface of narrow-line region clouds via photo-ionization from the quasar \citep{ster14}, and while this approach solves the problem of visibility of coronal lines in type 2 quasars, it may not explain the dependence on line width. Therefore, we suggest that instead of circumnuclear photo-ionization in type 2 quasars some or most of the coronal lines are due to shocks, possibly those produced when the radiatively-accelerated wind from the quasar runs into dense clouds. 

We draw inspiration from the observations by \citet{mazz13} who find that the extended ($\la 170$ pc) coronal line emission in NGC 1068 is likely dominated by shock excitation. These authors find that coronal line ratios are more easily explained by shock models than by photo-ionization models. Furthermore, they find that the kinematic structure of coronal lines is similar to that of several lower ionization features. Most intriguingly, the spatial and kinematic distribution of coronal lines is similar to that of very low ionization lines, such as [FeII], characteristic of shocks propagating into partly ionized medium. The latter point is particularly interesting as we observe the simultaneous rise of coronal lines and [SII] and [NI] in our sources. 

In NGC 1068, coronal lines are co-spatial with the radio jet, and thus \citet{mazz13} suggest that jet-driven shocks may play a role in line excitation. This is also a possibility for our sources and for type 1 quasars \citep{huse13,mull13}, especially in light of the fact that radio emission is strongly correlated with line kinematics (Section \ref{sec:radio}). However, we do not favor this scenario. While radio luminosity increases with [OIII] widths, and so do the fluxes of [FeVII], [SII] and [NI], there is no evidence that the latter lines increase in the composites made in bins of radio luminosity. What this suggests to us is that the primary driver for all these correlations is not the power of the radio emission, but the kinematics of the gas. Higher outflow velocity implies higher velocities of shocks driven into dense clouds, which enhances coronal emission, while the radio emission follows as discussed in Section \ref{sec:origin}. 

What sort of conditions would be required to produce coronal lines by driving low-density wind into high-density clouds? The typical shock velocities inside the shocked clouds implied by the observed coronal line ratios in NGC 1068 are $v_{\rm shock}=300-1000$ km/sec for cloud densities $n_{\rm cloud}=100-300$ cm$^{-3}$ \citep{mazz13}. To produce the same values in our picture, the incident wind velocity must be even higher, $v_{\rm wind}\sim v_{\rm shock}\sqrt{n_{\rm cloud}/n_{\rm wind}}$, where $n_{\rm wind}$ is the much lower density of the volume-filling component of the wind, so the required $v_{\rm wind}$ can be several thousand km/sec. Such high velocities are not achievable at large distances in low-luminosity active galaxies, where the radiatively-driven wind would get quenched by the interaction with the interstellar medium, but could be typical of quasar-driven winds \citep{moe09, zubo12, fauc12b} even on galaxy-wide scales \citep{arav13, liu13b}. 

\section{Discussion}
\label{sec:discussion}

\subsection{The origin of the line kinematics / radio / infrared relationships}
\label{sec:origin}

The tantalizing relationship between radio and infrared luminosities and the [OIII] line kinematics may contain interesting clues about the structure and the driving mechanism of ionized gas outflows in quasars, but its interpretation is not straightforward. For quite some time, it was thought that the extended ionized gas around active galaxies was much more likely to be found around radio-loud objects than around radio-quiet ones \citep{stoc87}. This was in line with theoretical ideas: powerful relativistic jets can inflate over-pressured cocoons which in turn sweep up galactic and inter-galactic medium plausibly resulting in narrow-line emission \citep{bege89}. 

With more sensitive observations, extended ionized gas emission has now been detected around radio-loud and radio-quiet quasars alike, indicating that gas nebulae around both types have roughly similar sizes \citep{liu13a, liu13b, liu14}, which reach 10 kkpc or greater at the highest luminosities \citep{gree11, liu13a, hain13, hain14, harr14}. The major difference appears to be that the nebulae around radio-loud objects tend to be more disturbed (lumpy) or more elongated, whereas the nebulae around radio-quiet objects are smooth and featureless \citep{liu13a}. The gas kinematics are disturbed and show outflow signatures not only in the central brightest parts of the nebulae, but all the way out to several kpc from the nucleus \citep{gree11, liu13b}. 

The recent discovery of the relationship between gas kinematics and radio emission within the radio-quiet population goes to the heart of the origin of the outflows in the majority of quasars, as well as of the origin of radio-quiet emission itself. Within the last year, there have been several new results on these issues \citep{cond13, huse13, mull13}. In this section we discuss pros and cons of various mechanisms proposed to explain the radio-quiet emission of quasars and the correlation between radio emission and gas kinematics. 

The observed gas kinematics / radio correlation is reminiscent of the classical work on narrow-line kinematics by \citet{veil91a, veil91b, veil91c} who found a correlation between [OIII]$\lambda$5007 line width and radio luminosity within a sample of 16 nearby Seyfert galaxies. In these and other nearby objects, high-resolution radio observations sometimes reveal a strong relationship between the position angle and morphology of the radio emission and the ionized gas emission, especially for the kinematically disturbed gas component (e.g., \citealt{veil91c, bowe95, cape97, falk98, cape99}). In some cases the jet appears to be at least in part responsible for the physical conditions in the narrow-line region (e.g., \citealt{krae98}). Interestingly, \citet{leip06} find a correlation between the size of the narrow-line region and the size of the radio source in nearby radio-quiet type 1 quasars. Thus, we first consider the possibility that the gas kinematics / radio relationship is established via direct interactions between the jet and the gas. 

In sources where the jet and the ionized gas emission are aligned, it is tempting to postulate that radio jets are either driving the outflow or provide shock-ionization, making them responsible for the morphology and / or excitation of the ionized gas, which may provide some basis for the radio / kinematics correlation. However, there may be a certain bias in reporting only those objects where high-resolution radio observations reveal a jet, especially if its position angle is not far off the orientation of the line emission. Far from every source shows evidence of jets in high-resolution radio observations \citep{veil91c, ho01}. In large unbiased samples the alignment between jets and ionized gas is statistically weak at best \citep{priv08}, although some authors find a stronger relationship \citep{schm03}, especially in compact radio sources \citep{shih13}. This remains a difficult issue because the morphology and the position angle of the radio emission and the ionized gas emission are a function of scale, and it is unclear which scales are relevant. Additionally, in principle the direction of photo-ionization by the active nucleus (which often determines the orientation of the nebula) and and the jet (which may be responsible for the disturbed kinematics) can be mis-aligned, further complicating the picture.  

We examined 20 objects from our sample that are located in Stripe 82 and were mapped by the VLA by \citet{hodg11}. These observations are deeper than FIRST (rms$=52\mu$Jy/beam vs 130$\mu$Jy/beam) and have higher spatial resolution (FWHM$=1.8$\arcsec\ vs 5\arcsec), better matched to the sizes of the quasar host galaxies at the typical redshifts of our sample (1\arcsec\ at $z=0.5$ is 6.1 kpc). Of the 20 sources, one object is a giant FRII radio galaxy in both datasets; 13 are point sources both in FIRST and in Stripe 82; and the remaining 7 are not detected in FIRST, but are all detected in Stripe 82 as point sources with fluxes between 0.4-0.7 mJy, i.e., just a factor of a couple below the catalog limit of the FIRST survey. None of the 19 objects showed resolved morphology in the higher resolution observations. Among type 2 quasars, \citet{lal10} find a mix of unresolved and slightly resolved sources at 0.8\arcsec\ resolution, but only a small fraction ($\sim 1/7$) with linear jet-like signatures. The radio emission of the $\sim 10$ kpc `super-bubble' quasar outflow \citep{gree12} is compact ($<1$ kpc, J.Wrobel, private communication). Overall we find it unlikely that large-scale ($\ga 5$ kpc) jets are responsible for the correlation between gas kinematics and radio emission we see in our sample. 

A more promising hypothesis is that in the radio-quiet objects the radio emission is due to a compact jet ($\la 1$ kpc) not resolved in the current radio observations which is driving the outflow of ionized gas \citep{spoo09, huse13, mull13}. Because the kinematics of the gas is disturbed on galaxy-wide scales (out to $\sim 10$ kpc, \citealt{liu13b}), the effect of the jet on the gas would need to be indirect in this scenario: the jet would need to inject the energy into the interstellar medium of the galaxy on small unresolved scales, with the energy then converted into a wide-angle outflow that engulfs the entire galaxy as observed. Although the observational test of this hypothesis seems straightforward (every radio-quiet quasar should have a jet), in practice testing this paradigm is difficult, in part because it is not clear that every extended radio structure represents a jet. Furthermore, radio emission of radio-quiet active galaxies tends to be compact \citep{naga03}, so it is often necessary to zoom in all the way into the central several pc in order to resolve the radio emission \citep{naga05}. On those scales, even when a linear structure seems unambiguously jet-like, it is not clear that it is sufficient to power an energetic galaxy-wide outflow. 

A completely different approach is followed by \citet{cond13} who argue, on the basis of the shape of the radio luminosity function, that the radio emission in radio-quiet quasars is mostly or entirely due to star formation in their host galaxies. In our objects, in favor of this hypothesis is the strong correlation between the infrared and the radio fluxes, which furthermore appear to lie on the extension of the locus of normal star-forming galaxies (Figure \ref{pic_wise}). However, at the luminosity of the radio-quiet subsample of $\log(\nu L_{\nu}[1.4{\rm GHz}], {\rm erg/sec})=40.0\pm 0.7$ (mean$=$median and standard deviation), the median star formation rate suggested by the classical radio / star-formation correlation \citep{helo85, bell03} would be an astonishing 400 $M_{\odot}$/year. Although the star formation rates of type 2 quasar hosts are among the highest in the population of active galaxies, in the ballpark of a few tens $M_{\odot}$/year \citep{zaka08}, they fall short of the values required to explain the observed radio luminosity, as was also noticed by \citet{lal10}. Furthermore, as was already noted above, the mid-infrared fluxes of our objects shown in Figure \ref{pic_wise} are dominated by the emission of the quasar-heated dust. Thus both radio luminosity and mid-infrared luminosity in Figure \ref{pic_wise} are unlikely to be due to star formation. 

Although compact jets cannot be ruled out by the existing data, we put forward an alternative suggestion for the origin of the radio-quiet emission. We propose that luminous quasars radiatively accelerate winds \citep{murr95, prog00} which then slam into the surrounding medium and drive shocks into the host galaxy. This shock blast initiated by the quasar-driven wind accelerates particles, just like what happens in a supernova remnant, which then produce synchrotron emission \citep{stoc92, jian10, fauc12b, zubo12}. To estimate the energetics of the wind required in this scenario, we assume for the moment that the efficiency of converting the kinetic energy of the outflow $L_{\rm wind}$ into radio synchrotron emission is similar in starburst-driven and quasar-driven winds. In a galaxy forming $\psi$ $M_{\odot}$/year worth of stars, the kinetic energy of the starburst-driven wind is $7\times 10^{41}\psi$ erg/sec \citep{leit99}. The same galaxy produces $\nu L_{\nu}$[1.4GHz]$=2.5\times 10^{37}\psi$ erg/sec worth of radio emission \citep{bell03}, with efficiency of $3.6\times 10^{-5}$ of converting kinetic energy into radio luminosity. 

Applying the same efficiency to our objects, we find that to reproduce our median radio luminosity of $10^{40}$ erg/sec we need $L_{\rm wind}=3\times 10^{44}$ erg/sec. The bolometric luminosities of our objects are poorly known, but using scalings presented by \citet{liu09} we can estimate that for the median [OIII] luminosity of our sample ($\log L$[OIII]/$L_{\odot}=8.9$) the bolometric luminosity is $8\times 10^{45}$ erg/sec. Thus the median fraction of the bolometric luminosity converted to the kinetic luminosity of the wind is 4\%, in rough agreement with our previous estimates based on observations of spatially resolved winds \citep{liu13b}. These ideas are further supported by our observation that $\nu L_{\nu}$[1.4GHz] has a close-to-quadratic dependence on [OIII] width (Figure \ref{pic_radio}), suggesting that the relationship between the kinetic energy of the ionized gas and radio luminosity is linear. 

In such heavily obscured quasars we cannot directly investigate the effect of the optical luminosity or even X-ray luminosity on gas kinematics or radio emission, which is an interesting direction to pursue using samples of type 1 quasars. In these objects, there is a known correlation between radio luminosity and their accretion power derived from ultra-violet luminosity \citep{falk95}. While this correlation is usually interpreted as a physical connection between accretion and resulting jets, an alternative explanation is that the accretion power leads to winds which in turn generate radio emission as described above without involving jets. 

\subsection{Radiatively-driven quasar winds}

\begin{figure}
\centering
\includegraphics[scale=0.4, clip=true, trim=0cm 0cm 0cm 0cm]{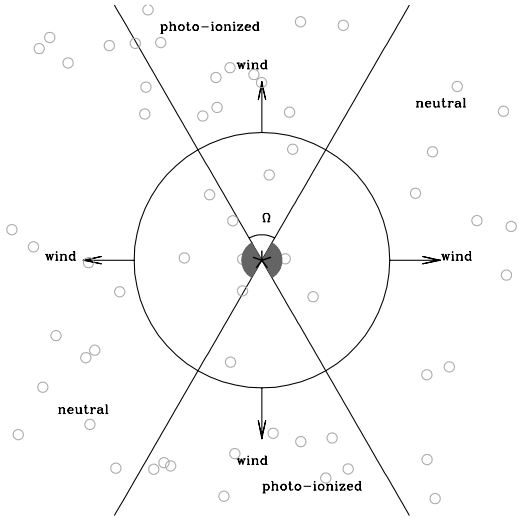}
\caption{A cartoon of our model for the narrow-line region of a quasar (central 5-point star). The quasar is surrounded by circumnuclear obscuring material (dark grey ``torus'') whose opening angle is $\Omega$. The wind propagates more isotropically than the ionizing radiation from the quasar which is concentrated in bi-cones marked ``photo-ionized''. Optical emission lines are produced by dense clouds (small circles). }
\label{pic_model}
\end{figure}

In Figure \ref{pic_model} we present a schematic of the components of the narrow-line regions of quasars discussed in this paper. The quasar is in the center and is characterized by its bolometric luminosity $L_{\rm bol}$. Obscuring material is distributed anisotropically around it (dark grey ``torus''), with $\Omega$ being the opening angle of the obscuration which determines the directions that can be photo-ionized by the direct quasar radiation. The relationship between the two values $L_{\rm bol}$ and $\Omega$ is not well known. Demographic studies of quasars continue to disagree on whether the obscured fraction stays constant or decreases as a function of luminosity \citep{trei08, lawr10}, but it is likely that at a given luminosity there is a wide distribution of $\Omega$ \citep{zaka06}. 

Another component of the model is the quasar-driven wind. The wind may be initially driven anisotropically, e.g., close to the equatorial plane of the accretion disk \citep{murr95, prog00}, and when it is first launched near the quasar, it is very fast, with velocities up to $\sim 0.1c$. The wind runs into the interstellar medium close to the nucleus and interacts with it, producing shock waves that propagate through the interstellar medium of the galaxy. If this medium is clumpy, the propagation of the wind proceeds along paths of least resistance, and the shape of the large-scale wind is determined largely by the distribution of the interstellar medium \citep{wagn13}, rather than by the initial anisotropies. Thus the wind is shown propagating in all directions, even those that are affected by the circumnuclear obscuration. 

The wind velocity $v_{\rm wind}$ likely varies as a function of the location in the galaxy and is not well-known observationally. In our data, the line-of-sight velocity dispersion of the optical emission lines is characteristic of the velocities of the clouds $v_0$ (light grey), not the velocity of the wind. The acceleration of clouds and their survival as they are impacted by the wind is a topic of active research \citep{mell02,coop09,aluz12,bara12,fauc12a}, but it seems natural that the velocity of the wind should be at least as high as the velocity of the clouds that it accelerates ($v_{\rm wind}\ga v_0$) and that they should be strongly related. Theoretical arguments suggest wind velocities of $v_{\rm wind} \ga 1000$ km/sec \citep{king05,zubo12,fauc12b}, in rough agreement with our inferred values of $v_0$. 

We do not observe $v_0$ directly; rather, we infer these values from the line-of-sight velocity width of the emission lines. The increase in shock-diagnostic lines [SII], [NI] and possibly [FeVII] as a function of [OIII] width provides evidence that the observed velocity width is a good proxy for the outflow velocity. Furthermore, the [OIII] width positively correlates with mid-infrared luminosity (Figure \ref{pic_wise}), which we take to suggest that the outflow is ultimately driven by the radiative pressure near the quasar \citep{murr95, prog00}. After the winds are launched, they interact with the interstellar medium in complex ways, which perhaps explains why the correlation between infrared luminosity and [OIII] velocity width (which our observations probe on galaxy-wide scales, \citealt{liu13b}) has a large scatter. 

\section{Conclusions}
\label{sec:conclusions}

\subsection{Gas kinematics and outflows}

In this paper we study the kinematics of ionized gas emission in 568 luminous obscured quasars from the sample of \citet{reye08}, primarily using their [OIII]$\lambda\lambda$4959,5007 emission lines. For every object, we determine a set of non-parametric measures of line asymmetry, velocity width and shape. We find that objects with blueshifted emission and line asymmetries are more prevalent in our sample, which we take as a signature that outlows in which the redshifted part is affected by dust extinction are common in our sample. The velocity widths we see in our sample (median $w_{80}=752$ km/sec and $w_{90}=1060$ km/sec, max $w_{80}=2918$ km/sec and $w_{90}=4780$ km/sec) are much higher than those in starburst galaxies (median $w_{90}\simeq 600$ km/sec, \citealt{rupk13a,hill14}).

Using the conversion between the line-of-sight velocity width of the line and the outflow velocity from Section \ref{sec:outflows}, we can estimate outflow velocities as $v_0\simeq w_{80}/1.5$. We find that for about half of the objects in our sample, $v_0$ ranges from $\sim$500 to $\sim$2000 km/sec. It is likely that in every object a range of cloud velocities is present; these estimates should be thought of as the median velocities in each source. From the observations presented here, it is not known which spatial scales dominate these estimates, but from our spatially resolved observations \citep{liu13b} it is clear that at least some of the clouds maintain similar velocities all the way out to $\sim 10$ kpc from the quasar. 

Neither the overall width nor the width of the narrower component in the multi-Gaussian line decomposition correlate with the velocity dispersion of the host galaxy. We therefore find no evidence for significant amounts of gas in dynamical equilibrium with the host (e.g., a component associated with the rotating galaxy disk). The higher velocity gas tends to be on average blueshifted by $\sim 100$ km/sec relative to the lower velocity gas. This may mean that the higher velocity gas is found on somewhat smaller spatial scales which are more prone to dust extinction. This conclusion is in line with the slight observed decrease of line-of-sight velocity dispersions of the gas as a function of the distance from the nucleus seen in spatially resolved observations of quasar winds \citep{liu13b}. The effect is very subtle (3\% decrease per each projected kpc), and overall high gas velocities are maintained over the entire host galaxy \citep{liu13b}. Other strong lines, such as [OII] and H$\beta$, also show outflow signatures, but significantly more mild than those seen in the [OIII] line. In the objects with the most extreme [OIII] kinematics, [OII] is much narrower than [OIII], with the kinematics of H$\beta$ being in between the two. 

The increase in shock-diagnostic lines [SII]$\lambda\lambda$4069,4076 and [NI]$\lambda\lambda$5198,5200 as a function of [OIII] width strong support for using the [OIII] width as a proxy for the outflow velocity. [OII] declines both with [OIII] width and [OIII] luminosity, suggesting that the [OII]-emitting regions are over-ionized by photo-ionization and quenched by increasing shock velocity. We find evidence for shock-ionization contribution to coronal lines [FeVII]. 

\subsection{Outflows and radio emission}

[OIII] velocity width correlates strongly with radio luminosity. We suggest that the radio emission is a by-product of the outflow activity, with particles accelerated on the shock fronts as the radiatively driven quasar wind propagates into the interstellar medium of the host galaxy (this mechanism was first hypothesized by \citealt{stoc92} to explain the low levels of radio emission in broad absorption line quasars). The median radio luminosity in our sample, $\nu L_{\nu}$[1.4GHz]$=10^{40}$ erg/sec, requires kinetic energy of the outflow of $L_{\rm wind}=3\times 10^{44}$ erg/sec if the efficiency of conversion in quasar-driven winds is similar to that in supernova-driven winds. The bolometric luminosities of our objects are poorly known, but we estimate that the median value for our sample is $8\times 10^{45}$ erg/sec and thus the kinetic luminosity of the wind is 4\% of the bolometric power. This idea is further reinforced by the approximately quadratic dependence $\nu L_{\nu}$[1.4GHz]$\propto$(width[OIII])$^2$, which implies a linear relationship between radio luminosity and kinetic energy of the outflow. [OIII] velocity is positively correlated with the infrared luminosity, which suggests that the ionized gas outflow is ultimately driven by the radiative pressure close to the accreting black hole. 

Another possible explanation for the [OIII] width / radio correlation is that the mechanical energy of the relativistic jet (which has not yet broken out of its host galaxy) is used to heat an overpressured cocoon \citep{bege89} which then launches the wind of ionized gas \citep{mull13}. As long as the jet is still compact, it appears as an unresolved radio core in FIRST observations. We use the scalings between the core radio luminosity and the jet kinetic energy by \citet{merl07} to estimate how much kinetic power would be required to produce the observed median radio luminosity. We find $L_{\rm jet}=3\times 10^{44}$ erg/sec, a value which is almost identical to the wind kinetic power obtained in the previous paragraph, even though they were estimated using completely different methods. This is likely not a coincidence: the efficiency of conversion of mechanical luminosity into synchrotron radiation is determined by the fraction of energy that can be converted into relativistic particles on shock fronts, which is likely independent of the origin of these shocks. 

The similarity of energy requirements underscores the difficulty of distinguishing between these two mechanisms. In the jet scenario one expects to see a jet in the high-resolution radio observations, whereas in the wind scenario radio emission is more diffuse and is present everywhere where shocks are propagating; however, in practice the collimated part of the jet does not dissipate energy very efficiently and so can be hard to detect. In both models, the morphology of the ionized gas depends much more strongly on the distribution of the interstellar medium than on the exact driving mechanism \citep{gaib11, wagn12, wagn13}. Thus, the morphology of ionized gas is not necessarily a useful clue. 

Radio spectral index could be a useful measurement for identifying recent or on-going particle acceleration (flatter synchrotron spectra for freshly accelerated, more energetic particles), but again particle acceleration may be happening in both scenarios. Interestingly, \citet{lal10} find flatter spectral indices in type 2 quasars than expected for jets in the standard geometry-based unification models, so a combination of spectral and morphological investigations in the radio may be worth pursuing further. 

One uncomfortable consequence of the jet scenario is that it requires for every radio-quiet quasar to have a powerful jet, with only a minority of them being active long enough to to break out of the galaxy (otherwise there would be too many extended radio sources). Another problem is that the jet scenario provides no ready explanation for the correlation between [OIII] width and infrared luminosity, unless jets are also directly connected to the accretion power \citep{falk95}.

In the wind scenario, if the initial distribution of matter is roughly spherically symmetric, so will be the ionized gas emission and the radio emission. In a disk galaxy, once the size of the wind reaches the vertical scaleheight of the disk it propagates largely along the path of least resistance perpendicular to the disk \citep{wagn13}, forming two symmetric bubbles on either side. Ionized gas emission is concentrated in shells and filaments, whereas the radio emission is filling the bubbles. In low-resolution data, one would see both the line emission and the radio emission oriented roughly in the same direction, so distinguishing this mechanism from jet-induced outflow requires high-quality observations. 

Such bubbles are directly seen in the radio in external galaxies \citep{ceci01, hota06} and in our own Milky Way, where the bubbles are known as ``the microwave haze'' \citep{fink04,su10}. These structures are also seen in X-rays which often closely trace the morphology of the radio emission and / or the ionized gas filaments \citep{ceci01, cros08, wang09}. In our Galaxy not only X-rays are observed \citep{sofu00}, but the structures as seen in gamma-rays and are known as ``the Fermi bubbles'' \citep{su10}. For comparison, the mechanical luminosity necessary to inflate the bubbles in NGC 3079 \citep{ceci01} is $\sim 30$ times lower than our median $L_{\rm wind}$, whereas that inferred for ``the Fermi bubbles'' in the Milky Way is $10^4-10^5$ times lower than our $L_{\rm wind}$. It is interesting that all of the examples above host jets \citep{ceci01, cros08, wang09, su12}, but it remains unclear whether this is universally true and whether they are contributing most of the required power. 

\subsection{Feedback in low- and high-luminosity active galaxies}

Our investigation focuses on the most luminous type 2 quasars at $z\la 1$. In these sources, we find strong evidence for quasar-driven winds on galaxy-wide scales, for radio emission associated with these winds, and for ionized gas velocities in excess of escape velocity from the galaxy. In lower luminosity active galaxies, many previous studies demonstrated that ionized gas is in dynamical equilibrium with the host galaxy and that radio emission is likely a bi-product of star-formation processes. Thus quasar-driven feedback may be present above some threshold in luminosity and absent below this threshold. We estimate this threshold by observing that in Figure \ref{pic_wise}, our sources separate well from star-forming galaxies at about $\nu L_{\nu}$[12\micron]$=2\times 10^{44}$ erg/sec. Using bolometric corrections from \citet{rich06}, \citet{liu09} and \citet{liu13b}, we estimate the corresponding bolometric luminosity to be $L_{\rm bol}=3\times 10^{45}$ erg/sec (uncertainty $\pm 0.4$ dex). 

Given the crudeness of this estimate, we consider it to be consistent with the value of $L_{\rm bol}=2.4\times 10^{45}$ erg/sec (uncertainty $\pm 0.3$ dex) suggested by \citet{veil13} based on incidence of molecular outflows in ultraluminous infrared galaxies \citep{veil13}. In principle the threshold value should be dependent on the depth of the potential (and therefore the stellar velocity dispersion), as well as the amount of gas that needs to be accelerated \citep{zubo12}, and it will be interesting to see whether these ideas are borne out in future analyses.  

\acknowledgments

NLZ would like to acknowledge useful conversations with M.J.Collinge (who suggested the diffuse nature of radio emission) and S.Tremaine (who suggested the analogy with Fermi bubbles), as well as with B.Groves, J.Krolik, D.Kushnir and J.Ostriker. The authors are grateful to the anonymous referee, H.Falcke, C.Norman, G.Richards, H.Spoon, J.Stern, J.Stocke, S. van Velzen and S.Veilleux for constructive comments during the review process. NLZ is thankful for the continued hospitality of the Institute for Advanced Study (Princeton) where part of this work was performed. 


\bibliographystyle{apj}
\bibliography{master}

\begin{thebibliography}{157}
\expandafter\ifx\csname natexlab\endcsname\relax\def\natexlab#1{#1}\fi

\bibitem[{{Aalto} {et~al.}(2012){Aalto}, {Garcia-Burillo}, {Muller}, {Winters},
  {van der Werf}, {Henkel}, {Costagliola}, \& {Neri}}]{aalt12}
{Aalto}, S., {Garcia-Burillo}, S., {Muller}, S., {et~al.} 2012, \aap, 537, A44

\bibitem[{{Al{\= u}zas} {et~al.}(2012){Al{\= u}zas}, {Pittard}, {Hartquist},
  {Falle}, \& {Langton}}]{aluz12}
{Al{\= u}zas}, R., {Pittard}, J.~M., {Hartquist}, T.~W., {Falle}, S.~A.~E.~G.,
  \& {Langton}, R. 2012, \mnras, 425, 2212

\bibitem[{{Alexander} {et~al.}(2010){Alexander}, {Swinbank}, {Smail},
  {McDermid}, \& {Nesvadba}}]{alex10}
{Alexander}, D.~M., {Swinbank}, A.~M., {Smail}, I., {McDermid}, R., \&
  {Nesvadba}, N.~P.~H. 2010, \mnras, 402, 2211

\bibitem[{{Antonucci} \& {Miller}(1985)}]{anto85}
{Antonucci}, R.~R.~J., \& {Miller}, J.~S. 1985, \apj, 297, 621

\bibitem[{{Arav} {et~al.}(2013){Arav}, {Borguet}, {Chamberlain}, {Edmonds}, \&
  {Danforth}}]{arav13}
{Arav}, N., {Borguet}, B., {Chamberlain}, C., {Edmonds}, D., \& {Danforth}, C.
  2013, \mnras, 436, 3286

\bibitem[{{Arav} {et~al.}(2008){Arav}, {Moe}, {Costantini}, {Korista}, {Benn},
  \& {Ellison}}]{arav08}
{Arav}, N., {Moe}, M., {Costantini}, E., {et~al.} 2008, \apj, 681, 954

\bibitem[{{Barai} {et~al.}(2012){Barai}, {Proga}, \& {Nagamine}}]{bara12}
{Barai}, P., {Proga}, D., \& {Nagamine}, K. 2012, \mnras, 424, 728

\bibitem[{{Barrows} {et~al.}(2013){Barrows}, {Sandberg Lacy}, {Kennefick},
  {Comerford}, {Kennefick}, \& {Berrier}}]{barr13}
{Barrows}, R.~S., {Sandberg Lacy}, C.~H., {Kennefick}, J., {et~al.} 2013, \apj,
  769, 95

\bibitem[{{Becker} {et~al.}(1995){Becker}, {White}, \& {Helfand}}]{beck95}
{Becker}, R.~H., {White}, R.~L., \& {Helfand}, D.~J. 1995, \apj, 450, 559

\bibitem[{{Begelman} \& {Cioffi}(1989)}]{bege89}
{Begelman}, M.~C., \& {Cioffi}, D.~F. 1989, \apjl, 345, L21

\bibitem[{{Bell}(2003)}]{bell03}
{Bell}, E.~F. 2003, \apj, 586, 794

\bibitem[{{Bennert} {et~al.}(2006){Bennert}, {Jungwiert}, {Komossa}, {Haas}, \&
  {Chini}}]{benn06}
{Bennert}, N., {Jungwiert}, B., {Komossa}, S., {Haas}, M., \& {Chini}, R. 2006,
  \aap, 456, 953

\bibitem[{{Blecha} {et~al.}(2013){Blecha}, {Loeb}, \& {Narayan}}]{blec13}
{Blecha}, L., {Loeb}, A., \& {Narayan}, R. 2013, \mnras, 429, 2594

\bibitem[{{Bower} {et~al.}(1995){Bower}, {Wilson}, {Morse}, {Gelderman},
  {Whittle}, \& {Mulchaey}}]{bowe95}
{Bower}, G., {Wilson}, A., {Morse}, J.~A., {et~al.} 1995, \apj, 454, 106

\bibitem[{{Brinchmann} {et~al.}(2008){Brinchmann}, {Kunth}, \&
  {Durret}}]{brin08}
{Brinchmann}, J., {Kunth}, D., \& {Durret}, F. 2008, \aap, 485, 657

\bibitem[{{Bruzual} \& {Charlot}(2003)}]{bruz03}
{Bruzual}, G., \& {Charlot}, S. 2003, \mnras, 344, 1000

\bibitem[{{Cano-D{\'{\i}}az} {et~al.}(2012){Cano-D{\'{\i}}az}, {Maiolino},
  {Marconi}, {Netzer}, {Shemmer}, \& {Cresci}}]{cano12}
{Cano-D{\'{\i}}az}, M., {Maiolino}, R., {Marconi}, A., {et~al.} 2012, \aap,
  537, L8

\bibitem[{{Capetti} {et~al.}(1997){Capetti}, {Axon}, \& {Macchetto}}]{cape97}
{Capetti}, A., {Axon}, D.~J., \& {Macchetto}, F.~D. 1997, \apj, 487, 560

\bibitem[{{Capetti} {et~al.}(1999){Capetti}, {Axon}, {Macchetto}, {Marconi}, \&
  {Winge}}]{cape99}
{Capetti}, A., {Axon}, D.~J., {Macchetto}, F.~D., {Marconi}, A., \& {Winge}, C.
  1999, \apj, 516, 187

\bibitem[{{Cecil} {et~al.}(2001){Cecil}, {Bland-Hawthorn}, {Veilleux}, \&
  {Filippenko}}]{ceci01}
{Cecil}, G., {Bland-Hawthorn}, J., {Veilleux}, S., \& {Filippenko}, A.~V. 2001,
  \apj, 555, 338

\bibitem[{{Cicone} {et~al.}(2014){Cicone}, {Maiolino}, {Sturm},
  {Graci{\'a}-Carpio}, {Feruglio}, {Neri}, {Aalto}, {Davies}, {Fiore},
  {Fischer}, {Garc{\'{\i}}a-Burillo}, {Gonz{\'a}lez-Alfonso},
  {Hailey-Dunsheath}, {Piconcelli}, \& {Veilleux}}]{cico14}
{Cicone}, C., {Maiolino}, R., {Sturm}, E., {et~al.} 2014, \aap, 562, A21

\bibitem[{{Comerford} {et~al.}(2009)}]{come09a}
{Comerford}, J.~M., {et~al.} 2009, \apj, 698, 956

\bibitem[{{Condon} {et~al.}(1998){Condon}, {Cotton}, {Greisen}, {Yin},
  {Perley}, {Taylor}, \& {Broderick}}]{cond98}
{Condon}, J.~J., {Cotton}, W.~D., {Greisen}, E.~W., {et~al.} 1998, \aj, 115,
  1693

\bibitem[{{Condon} {et~al.}(2013){Condon}, {Kellermann}, {Kimball},
  {Ivezi{\'c}}, \& {Perley}}]{cond13}
{Condon}, J.~J., {Kellermann}, K.~I., {Kimball}, A.~E., {Ivezi{\'c}}, {\v Z}.,
  \& {Perley}, R.~A. 2013, \apj, 768, 37

\bibitem[{{Cooper} {et~al.}(2009){Cooper}, {Bicknell}, {Sutherland}, \&
  {Bland-Hawthorn}}]{coop09}
{Cooper}, J.~L., {Bicknell}, G.~V., {Sutherland}, R.~S., \& {Bland-Hawthorn},
  J. 2009, \apj, 703, 330

\bibitem[{{Crenshaw} \& {Kraemer}(2000)}]{cren00}
{Crenshaw}, D.~M., \& {Kraemer}, S.~B. 2000, \apjl, 532, L101

\bibitem[{{Crenshaw} {et~al.}(2003){Crenshaw}, {Kraemer}, \& {George}}]{cren03}
{Crenshaw}, D.~M., {Kraemer}, S.~B., \& {George}, I.~M. 2003, \araa, 41, 117

\bibitem[{{Croston} {et~al.}(2008){Croston}, {Hardcastle}, {Kharb}, {Kraft}, \&
  {Hota}}]{cros08}
{Croston}, J.~H., {Hardcastle}, M.~J., {Kharb}, P., {Kraft}, R.~P., \& {Hota},
  A. 2008, \apj, 688, 190

\bibitem[{{De Robertis} \& {Osterbrock}(1984)}]{dero84}
{De Robertis}, M.~M., \& {Osterbrock}, D.~E. 1984, \apj, 286, 171

\bibitem[{{Dunn} {et~al.}(2010){Dunn}, {Bautista}, {Arav}, {Moe}, {Korista},
  {Costantini}, {Benn}, {Ellison}, \& {Edmonds}}]{dunn10}
{Dunn}, J.~P., {Bautista}, M., {Arav}, N., {et~al.} 2010, \apj, 709, 611

\bibitem[{{Fabian}(2012)}]{fabi12}
{Fabian}, A.~C. 2012, \araa, 50, 455

\bibitem[{{Falcke} {et~al.}(1995){Falcke}, {Malkan}, \& {Biermann}}]{falk95}
{Falcke}, H., {Malkan}, M.~A., \& {Biermann}, P.~L. 1995, \aap, 298, 375

\bibitem[{{Falcke} {et~al.}(1998){Falcke}, {Wilson}, \& {Simpson}}]{falk98}
{Falcke}, H., {Wilson}, A.~S., \& {Simpson}, C. 1998, \apj, 502, 199

\bibitem[{{Faucher-Gigu{\`e}re} \& {Quataert}(2012)}]{fauc12b}
{Faucher-Gigu{\`e}re}, C.-A., \& {Quataert}, E. 2012, \mnras, 425, 605

\bibitem[{{Faucher-Gigu{\`e}re} {et~al.}(2012){Faucher-Gigu{\`e}re},
  {Quataert}, \& {Murray}}]{fauc12a}
{Faucher-Gigu{\`e}re}, C.-A., {Quataert}, E., \& {Murray}, N. 2012, \mnras,
  420, 1347

\bibitem[{{Feruglio} {et~al.}(2010){Feruglio}, {Maiolino}, {Piconcelli},
  {Menci}, {Aussel}, {Lamastra}, \& {Fiore}}]{feru10}
{Feruglio}, C., {Maiolino}, R., {Piconcelli}, E., {et~al.} 2010, \aap, 518,
  L155

\bibitem[{{Finkbeiner}(2004)}]{fink04}
{Finkbeiner}, D.~P. 2004, \apj, 614, 186

\bibitem[{{Fischer} {et~al.}(2010){Fischer}, {Crenshaw}, {Kraemer}, {Schmitt},
  \& {Trippe}}]{fisc10}
{Fischer}, T.~C., {Crenshaw}, D.~M., {Kraemer}, S.~B., {Schmitt}, H.~R., \&
  {Trippe}, M.~L. 2010, \aj, 140, 577

\bibitem[{{Fu} {et~al.}(2012){Fu}, {Yan}, {Myers}, {Stockton}, {Djorgovski},
  {Aldering}, \& {Rich}}]{fu12}
{Fu}, H., {Yan}, L., {Myers}, A.~D., {et~al.} 2012, \apj, 745, 67

\bibitem[{{Gaibler} {et~al.}(2011){Gaibler}, {Khochfar}, \& {Krause}}]{gaib11}
{Gaibler}, V., {Khochfar}, S., \& {Krause}, M. 2011, \mnras, 411, 155

\bibitem[{{Gelbord} {et~al.}(2009){Gelbord}, {Mullaney}, \& {Ward}}]{gelb09}
{Gelbord}, J.~M., {Mullaney}, J.~R., \& {Ward}, M.~J. 2009, \mnras, 397, 172

\bibitem[{{Grandi}(1978)}]{gran78}
{Grandi}, S.~A. 1978, \apj, 221, 501

\bibitem[{{Greene} \& {Ho}(2005)}]{gree05o3}
{Greene}, J.~E., \& {Ho}, L.~C. 2005, \apj, 627, 721

\bibitem[{{Greene} \& {Ho}(2006)}]{gree06}
---. 2006, \apj, 641, 117

\bibitem[{{Greene} {et~al.}(2011){Greene}, {Zakamska}, {Ho}, \&
  {Barth}}]{gree11}
{Greene}, J.~E., {Zakamska}, N.~L., {Ho}, L.~C., \& {Barth}, A.~J. 2011, \apj,
  732, 9

\bibitem[{{Greene} {et~al.}(2009){Greene}, {Zakamska}, {Liu}, {Barth}, \&
  {Ho}}]{gree09}
{Greene}, J.~E., {Zakamska}, N.~L., {Liu}, X., {Barth}, A.~J., \& {Ho}, L.~C.
  2009, \apj, 702, 441

\bibitem[{{Greene} {et~al.}(2012){Greene}, {Zakamska}, \& {Smith}}]{gree12}
{Greene}, J.~E., {Zakamska}, N.~L., \& {Smith}, P.~S. 2012, \apj, 746, 86

\bibitem[{{Hainline} {et~al.}(2013){Hainline}, {Hickox}, {Greene}, {Myers}, \&
  {Zakamska}}]{hain13}
{Hainline}, K.~N., {Hickox}, R., {Greene}, J.~E., {Myers}, A.~D., \&
  {Zakamska}, N.~L. 2013, \apj, 774, 145

\bibitem[{{Hainline} {et~al.}(2014){Hainline}, {Hickox}, {Greene}, {Myers},
  {Zakamska}, {Liu}, \& {Liu}}]{hain14}
{Hainline}, K.~N., {Hickox}, R.~C., {Greene}, J.~E., {et~al.} 2014, ArXiv
  e-prints

\bibitem[{{Harrison} {et~al.}(2014){Harrison}, {Alexander}, {Mullaney}, \&
  {Swinbank}}]{harr14}
{Harrison}, C.~M., {Alexander}, D.~M., {Mullaney}, J.~R., \& {Swinbank}, A.~M.
  2014, ArXiv e-prints

\bibitem[{{Harrison} {et~al.}(2012){Harrison}, {Alexander}, {Swinbank},
  {Smail}, {Alaghband-Zadeh}, {Bauer}, {Chapman}, {Del Moro}, {Hickox},
  {Ivison}, {Men{\'e}ndez-Delmestre}, {Mullaney}, \& {Nesvadba}}]{harr12}
{Harrison}, C.~M., {Alexander}, D.~M., {Swinbank}, A.~M., {et~al.} 2012,
  \mnras, 426, 1073

\bibitem[{{Heckman} {et~al.}(1981){Heckman}, {Miley}, {van Breugel}, \&
  {Butcher}}]{heck81}
{Heckman}, T.~M., {Miley}, G.~K., {van Breugel}, W.~J.~M., \& {Butcher}, H.~R.
  1981, \apj, 247, 403

\bibitem[{{Helou} {et~al.}(1985){Helou}, {Soifer}, \&
  {Rowan-Robinson}}]{helo85}
{Helou}, G., {Soifer}, B.~T., \& {Rowan-Robinson}, M. 1985, \apjl, 298, L7

\bibitem[{{Hill} \& {Zakamska}(2014)}]{hill14}
{Hill}, M.~J., \& {Zakamska}, N.~L. 2014, \mnras, 439, 2701

\bibitem[{{Ho} \& {Peng}(2001)}]{hopeng01}
{Ho}, L.~C., \& {Peng}, C.~Y. 2001, \apj, 555, 650

\bibitem[{{Ho} \& {Ulvestad}(2001)}]{ho01}
{Ho}, L.~C., \& {Ulvestad}, J.~S. 2001, \apjs, 133, 77

\bibitem[{{Hodge} {et~al.}(2011){Hodge}, {Becker}, {White}, {Richards}, \&
  {Zeimann}}]{hodg11}
{Hodge}, J.~A., {Becker}, R.~H., {White}, R.~L., {Richards}, G.~T., \&
  {Zeimann}, G.~R. 2011, \aj, 142, 3

\bibitem[{{Hopkins} {et~al.}(2006){Hopkins}, {Hernquist}, {Cox}, {Di Matteo},
  {Robertson}, \& {Springel}}]{hopk06}
{Hopkins}, P.~F., {Hernquist}, L., {Cox}, T.~J., {et~al.} 2006, \apjs, 163, 1

\bibitem[{{Hota} \& {Saikia}(2006)}]{hota06}
{Hota}, A., \& {Saikia}, D.~J. 2006, \mnras, 371, 945

\bibitem[{{Husemann} {et~al.}(2013){Husemann}, {Wisotzki}, {S{\'a}nchez}, \&
  {Jahnke}}]{huse13}
{Husemann}, B., {Wisotzki}, L., {S{\'a}nchez}, S.~F., \& {Jahnke}, K. 2013,
  \aap, 549, A43

\bibitem[{{Ivezi{\'c}} {et~al.}(2002){Ivezi{\'c}}, {Menou}, {Knapp}, {Strauss},
  {Lupton}, {Vanden Berk}, {Richards}, {Tremonti}, {Weinstein}, {Anderson},
  {Bahcall}, {Becker}, {Bernardi}, {Blanton}, {Eisenstein}, {Fan},
  {Finkbeiner}, {Finlator}, {Frieman}, {Gunn}, {Hall}, {Kim}, {Kinkhabwala},
  {Narayanan}, {Rockosi}, {Schlegel}, {Schneider}, {Strateva}, {SubbaRao},
  {Thakar}, {Voges}, {White}, {Yanny}, {Brinkmann}, {Doi}, {Fukugita},
  {Hennessy}, {Munn}, {Nichol}, \& {York}}]{ivez02}
{Ivezi{\'c}}, {\v Z}., {Menou}, K., {Knapp}, G.~R., {et~al.} 2002, \aj, 124,
  2364

\bibitem[{{Jia} {et~al.}(2013){Jia}, {Ptak}, {Heckman}, \& {Zakamska}}]{jia13}
{Jia}, J., {Ptak}, A., {Heckman}, T., \& {Zakamska}, N.~L. 2013, \apj, 777, 27

\bibitem[{{Jiang} {et~al.}(2007){Jiang}, {Fan}, {Ivezi{\'c}}, {Richards},
  {Schneider}, {Strauss}, \& {Kelly}}]{jian07a}
{Jiang}, L., {Fan}, X., {Ivezi{\'c}}, {\v Z}., {et~al.} 2007, \apj, 656, 680

\bibitem[{{Jiang} {et~al.}(2010){Jiang}, {Ciotti}, {Ostriker}, \&
  {Spitkovsky}}]{jian10}
{Jiang}, Y.-F., {Ciotti}, L., {Ostriker}, J.~P., \& {Spitkovsky}, A. 2010,
  \apj, 711, 125

\bibitem[{{Kellermann} {et~al.}(1989){Kellermann}, {Sramek}, {Schmidt},
  {Shaffer}, \& {Green}}]{kell89}
{Kellermann}, K.~I., {Sramek}, R., {Schmidt}, M., {Shaffer}, D.~B., \& {Green},
  R. 1989, \aj, 98, 1195

\bibitem[{{Kim} {et~al.}(2006){Kim}, {Ho}, \& {Im}}]{kim06}
{Kim}, M., {Ho}, L.~C., \& {Im}, M. 2006, \apj, 642, 702

\bibitem[{{King}(2005)}]{king05}
{King}, A. 2005, \apjl, 635, L121

\bibitem[{{Komossa} {et~al.}(2008){Komossa}, {Zhou}, {Wang}, {Ajello}, {Ge},
  {Greiner}, {Lu}, {Salvato}, {Saxton}, {Shan}, {Xu}, \& {Yuan}}]{komo08}
{Komossa}, S., {Zhou}, H., {Wang}, T., {et~al.} 2008, \apjl, 678, L13

\bibitem[{{Kraemer} {et~al.}(1998){Kraemer}, {Ruiz}, \& {Crenshaw}}]{krae98}
{Kraemer}, S.~B., {Ruiz}, J.~R., \& {Crenshaw}, D.~M. 1998, \apj, 508, 232

\bibitem[{Kramida {et~al.}(2013)Kramida, {Yu.~Ralchenko}, Reader, \& {and NIST
  ASD Team}}]{kram13}
Kramida, A., {Yu.~Ralchenko}, Reader, J., \& {and NIST ASD Team}. 2013, {NIST
  Atomic Spectra Database (ver. 5.1), [Online]. Available:
  {\tt{http://physics.nist.gov/asd}} [2014, January 2]. National Institute of
  Standards and Technology, Gaithersburg, MD.}

\bibitem[{{Lacy} {et~al.}(2004){Lacy}, {Storrie-Lombardi}, {Sajina},
  {Appleton}, {Armus}, {Chapman}, {Choi}, {Fadda}, {Fang}, {Frayer},
  {Heinrichsen}, {Helou}, {Im}, {Marleau}, {Masci}, {Shupe}, {Soifer},
  {Surace}, {Teplitz}, {Wilson}, \& {Yan}}]{lacy04}
{Lacy}, M., {Storrie-Lombardi}, L.~J., {Sajina}, A., {et~al.} 2004, \apjs, 154,
  166

\bibitem[{{Lacy} {et~al.}(2013){Lacy}, {Ridgway}, {Gates}, {Nielsen}, {Petric},
  {Sajina}, {Urrutia}, {Cox Drews}, {Harrison}, {Seymour}, \&
  {Storrie-Lombardi}}]{lacy13}
{Lacy}, M., {Ridgway}, S.~E., {Gates}, E.~L., {et~al.} 2013, \apjs, 208, 24

\bibitem[{{Lal} \& {Ho}(2010)}]{lal10}
{Lal}, D.~V., \& {Ho}, L.~C. 2010, \aj, 139, 1089

\bibitem[{{LaMassa} {et~al.}(2011){LaMassa}, {Heckman}, {Ptak}, {Martins},
  {Wild}, {Sonnentrucker}, \& {Hornschemeier}}]{lama11}
{LaMassa}, S.~M., {Heckman}, T.~M., {Ptak}, A., {et~al.} 2011, \apj, 729, 52

\bibitem[{{Lawrence} \& {Elvis}(2010)}]{lawr10}
{Lawrence}, A., \& {Elvis}, M. 2010, \apj, 714, 561

\bibitem[{{Leipski} {et~al.}(2006){Leipski}, {Falcke}, {Bennert}, \&
  {H{\"u}ttemeister}}]{leip06}
{Leipski}, C., {Falcke}, H., {Bennert}, N., \& {H{\"u}ttemeister}, S. 2006,
  \aap, 455, 161

\bibitem[{{Leitherer} {et~al.}(1999){Leitherer}, {Schaerer}, {Goldader},
  {Gonz{\'a}lez Delgado}, {Robert}, {Kune}, {de Mello}, {Devost}, \&
  {Heckman}}]{leit99}
{Leitherer}, C., {Schaerer}, D., {Goldader}, J.~D., {et~al.} 1999, \apjs, 123,
  3

\bibitem[{{Liu} {et~al.}(2014){Liu}, {Zakamska}, \& {Greene}}]{liu14}
{Liu}, G., {Zakamska}, N.~L., \& {Greene}, J.~E. 2014, ArXiv e-prints

\bibitem[{{Liu} {et~al.}(2013{\natexlab{a}}){Liu}, {Zakamska}, {Greene},
  {Nesvadba}, \& {Liu}}]{liu13a}
{Liu}, G., {Zakamska}, N.~L., {Greene}, J.~E., {Nesvadba}, N.~P.~H., \& {Liu},
  X. 2013{\natexlab{a}}, \mnras, 430, 2327

\bibitem[{{Liu} {et~al.}(2013{\natexlab{b}}){Liu}, {Zakamska}, {Greene},
  {Nesvadba}, \& {Liu}}]{liu13b}
---. 2013{\natexlab{b}}, \mnras, 436, 2576

\bibitem[{{Liu} {et~al.}(2010){Liu}, {Shen}, {Strauss}, \& {Greene}}]{liu10a}
{Liu}, X., {Shen}, Y., {Strauss}, M.~A., \& {Greene}, J.~E. 2010, \apj, 708,
  427

\bibitem[{{Liu} {et~al.}(2009){Liu}, {Zakamska}, {Greene}, {Strauss}, {Krolik},
  \& {Heckman}}]{liu09}
{Liu}, X., {Zakamska}, N.~L., {Greene}, J.~E., {et~al.} 2009, \apj, 702, 1098

\bibitem[{{Mazzalay} {et~al.}(2013){Mazzalay}, {Rodr{\'{\i}}guez-Ardila},
  {Komossa}, \& {McGregor}}]{mazz13}
{Mazzalay}, X., {Rodr{\'{\i}}guez-Ardila}, A., {Komossa}, S., \& {McGregor},
  P.~J. 2013, \mnras, 430, 2411

\bibitem[{{Mellema} {et~al.}(2002){Mellema}, {Kurk}, \&
  {R{\"o}ttgering}}]{mell02}
{Mellema}, G., {Kurk}, J.~D., \& {R{\"o}ttgering}, H.~J.~A. 2002, \aap, 395,
  L13

\bibitem[{{Merloni} \& {Heinz}(2007)}]{merl07}
{Merloni}, A., \& {Heinz}, S. 2007, \mnras, 381, 589

\bibitem[{{Moe} {et~al.}(2009){Moe}, {Arav}, {Bautista}, \& {Korista}}]{moe09}
{Moe}, M., {Arav}, N., {Bautista}, M.~A., \& {Korista}, K.~T. 2009, \apj, 706,
  525

\bibitem[{{Monreal-Ibero} {et~al.}(2013){Monreal-Ibero}, {Walsh},
  {Westmoquette}, \& {V{\'{\i}}lchez}}]{monr13}
{Monreal-Ibero}, A., {Walsh}, J.~R., {Westmoquette}, M.~S., \&
  {V{\'{\i}}lchez}, J.~M. 2013, \aap, 553, A57

\bibitem[{{Morton}(1991)}]{mort91}
{Morton}, D.~C. 1991, \apjs, 77, 119

\bibitem[{{Moy} \& {Rocca-Volmerange}(2002)}]{moy02}
{Moy}, E., \& {Rocca-Volmerange}, B. 2002, \aap, 383, 46

\bibitem[{{Mullaney} {et~al.}(2013){Mullaney}, {Alexander}, {Fine}, {Goulding},
  {Harrison}, \& {Hickox}}]{mull13}
{Mullaney}, J.~R., {Alexander}, D.~M., {Fine}, S., {et~al.} 2013, \mnras, 433,
  622

\bibitem[{{Mullaney} {et~al.}(2009){Mullaney}, {Ward}, {Done}, {Ferland}, \&
  {Schurch}}]{mull09}
{Mullaney}, J.~R., {Ward}, M.~J., {Done}, C., {Ferland}, G.~J., \& {Schurch},
  N. 2009, \mnras, 394, L16

\bibitem[{{M{\"u}ller-S{\'a}nchez} {et~al.}(2011){M{\"u}ller-S{\'a}nchez},
  {Prieto}, {Hicks}, {Vives-Arias}, {Davies}, {Malkan}, {Tacconi}, \&
  {Genzel}}]{mull11}
{M{\"u}ller-S{\'a}nchez}, F., {Prieto}, M.~A., {Hicks}, E.~K.~S., {et~al.}
  2011, \apj, 739, 69

\bibitem[{{Murray} {et~al.}(1995){Murray}, {Chiang}, {Grossman}, \&
  {Voit}}]{murr95}
{Murray}, N., {Chiang}, J., {Grossman}, S.~A., \& {Voit}, G.~M. 1995, \apj,
  451, 498

\bibitem[{{Nagar} {et~al.}(2005){Nagar}, {Falcke}, \& {Wilson}}]{naga05}
{Nagar}, N.~M., {Falcke}, H., \& {Wilson}, A.~S. 2005, \aap, 435, 521

\bibitem[{{Nagar} {et~al.}(2003){Nagar}, {Wilson}, {Falcke}, {Veilleux}, \&
  {Maiolino}}]{naga03}
{Nagar}, N.~M., {Wilson}, A.~S., {Falcke}, H., {Veilleux}, S., \& {Maiolino},
  R. 2003, \aap, 409, 115

\bibitem[{{Nelson} \& {Whittle}(1996)}]{nels96}
{Nelson}, C.~H., \& {Whittle}, M. 1996, \apj, 465, 96

\bibitem[{{Nesvadba} {et~al.}(2008){Nesvadba}, {Lehnert}, {De Breuck},
  {Gilbert}, \& {van Breugel}}]{nesv08}
{Nesvadba}, N.~P.~H., {Lehnert}, M.~D., {De Breuck}, C., {Gilbert}, A.~M., \&
  {van Breugel}, W. 2008, \aap, 491, 407

\bibitem[{{Nesvadba} {et~al.}(2006){Nesvadba}, {Lehnert}, {Eisenhauer},
  {Gilbert}, {Tecza}, \& {Abuter}}]{nesv06}
{Nesvadba}, N.~P.~H., {Lehnert}, M.~D., {Eisenhauer}, F., {et~al.} 2006, \apj,
  650, 693

\bibitem[{{Novak} {et~al.}(2011){Novak}, {Ostriker}, \& {Ciotti}}]{nova11}
{Novak}, G.~S., {Ostriker}, J.~P., \& {Ciotti}, L. 2011, \apj, 737, 26

\bibitem[{{Osterbrock} \& {Ferland}(2006)}]{oste06}
{Osterbrock}, D.~E., \& {Ferland}, G.~J. 2006, {Astrophysics of gaseous nebulae
  and active galactic nuclei} (Sausalito, CA: University Science Books)

\bibitem[{{Privon} {et~al.}(2008){Privon}, {O'Dea}, {Baum}, {Axon}, {Kharb},
  {Buchanan}, {Sparks}, \& {Chiaberge}}]{priv08}
{Privon}, G.~C., {O'Dea}, C.~P., {Baum}, S.~A., {et~al.} 2008, \apjs, 175, 423

\bibitem[{{Proga} {et~al.}(2000){Proga}, {Stone}, \& {Kallman}}]{prog00}
{Proga}, D., {Stone}, J.~M., \& {Kallman}, T.~R. 2000, \apj, 543, 686

\bibitem[{{Ptak} {et~al.}(2006){Ptak}, {Zakamska}, {Strauss}, {Krolik},
  {Heckman}, {Schneider}, \& {Brinkmann}}]{ptak06}
{Ptak}, A., {Zakamska}, N.~L., {Strauss}, M.~A., {et~al.} 2006, \apj, 637, 147

\bibitem[{{Reyes} {et~al.}(2008){Reyes}, {Zakamska}, {Strauss}, {Green},
  {Krolik}, {Shen}, {Richards}, {Anderson}, \& {Schneider}}]{reye08}
{Reyes}, R., {Zakamska}, N.~L., {Strauss}, M.~A., {et~al.} 2008, \aj, 136, 2373

\bibitem[{{Richards} {et~al.}(2006){Richards}, {Lacy}, {Storrie-Lombardi},
  {Hall}, {Gallagher}, {Hines}, {Fan}, {Papovich}, {Vanden Berk}, {Trammell},
  {Schneider}, {Vestergaard}, {York}, {Jester}, {Anderson}, {Budav{\'a}ri}, \&
  {Szalay}}]{rich06}
{Richards}, G.~T., {Lacy}, M., {Storrie-Lombardi}, L.~J., {et~al.} 2006, \apjs,
  166, 470

\bibitem[{{Rodr{\'{\i}}guez-Ardila} {et~al.}(2011){Rodr{\'{\i}}guez-Ardila},
  {Prieto}, {Portilla}, \& {Tejeiro}}]{rodr11}
{Rodr{\'{\i}}guez-Ardila}, A., {Prieto}, M.~A., {Portilla}, J.~G., \&
  {Tejeiro}, J.~M. 2011, \apj, 743, 100

\bibitem[{{Rosario} {et~al.}(2013){Rosario}, {Burtscher}, {Davies}, {Genzel},
  {Lutz}, \& {Tacconi}}]{rosa13}
{Rosario}, D.~J., {Burtscher}, L., {Davies}, R., {et~al.} 2013, \apj, 778, 94

\bibitem[{{Rupke} \& {Veilleux}(2011)}]{rupk11}
{Rupke}, D.~S.~N., \& {Veilleux}, S. 2011, \apjl, 729, L27

\bibitem[{{Rupke} \& {Veilleux}(2013{\natexlab{a}})}]{rupk13b}
---. 2013{\natexlab{a}}, \apjl, 775, L15

\bibitem[{{Rupke} \& {Veilleux}(2013{\natexlab{b}})}]{rupk13a}
---. 2013{\natexlab{b}}, \apj, 768, 75

\bibitem[{{Schaerer} {et~al.}(1999){Schaerer}, {Contini}, \& {Kunth}}]{scha99}
{Schaerer}, D., {Contini}, T., \& {Kunth}, D. 1999, \aap, 341, 399

\bibitem[{{Schmitt} {et~al.}(2003){Schmitt}, {Donley}, {Antonucci},
  {Hutchings}, {Kinney}, \& {Pringle}}]{schm03}
{Schmitt}, H.~R., {Donley}, J.~L., {Antonucci}, R.~R.~J., {et~al.} 2003, \apj,
  597, 768

\bibitem[{{Shen} {et~al.}(2010){Shen}, {Liu}, {Greene}, \& {Strauss}}]{shen11}
{Shen}, Y., {Liu}, X., {Greene}, J., \& {Strauss}, M. 2010, \apj, submitted
  (astroph/1011.5246)

\bibitem[{{Shih} {et~al.}(2013){Shih}, {Stockton}, \& {Kewley}}]{shih13}
{Shih}, H.-Y., {Stockton}, A., \& {Kewley}, L. 2013, \apj, 772, 138

\bibitem[{{Silk} \& {Rees}(1998)}]{silk98}
{Silk}, J., \& {Rees}, M.~J. 1998, \aap, 331, L1

\bibitem[{{Sofue}(2000)}]{sofu00}
{Sofue}, Y. 2000, \apj, 540, 224

\bibitem[{{Spoon} \& {Holt}(2009)}]{spoo09}
{Spoon}, H.~W.~W., \& {Holt}, J. 2009, \apjl, 702, L42

\bibitem[{{Springel} {et~al.}(2005){Springel}, {Di Matteo}, \&
  {Hernquist}}]{spri05}
{Springel}, V., {Di Matteo}, T., \& {Hernquist}, L. 2005, \mnras, 361, 776

\bibitem[{{Steinhardt} \& {Silverman}(2013)}]{stei13}
{Steinhardt}, C.~L., \& {Silverman}, J.~D. 2013, \pasj, 65, 82

\bibitem[{{Stern} {et~al.}(2005){Stern}, {Eisenhardt}, {Gorjian}, {Kochanek},
  {Caldwell}, {Eisenstein}, {Brodwin}, {Brown}, {Cool}, {Dey}, {Green},
  {Jannuzi}, {Murray}, {Pahre}, \& {Willner}}]{ster05}
{Stern}, D., {Eisenhardt}, P., {Gorjian}, V., {et~al.} 2005, \apj, 631, 163

\bibitem[{{Stern} \& {Laor}(2012)}]{ster12}
{Stern}, J., \& {Laor}, A. 2012, \mnras, 426, 2703

\bibitem[{{Stern} {et~al.}(2014){Stern}, {Laor}, \& {Baskin}}]{ster14}
{Stern}, J., {Laor}, A., \& {Baskin}, A. 2014, \mnras, 438, 901

\bibitem[{{Stocke} {et~al.}(1992){Stocke}, {Morris}, {Weymann}, \&
  {Foltz}}]{stoc92}
{Stocke}, J.~T., {Morris}, S.~L., {Weymann}, R.~J., \& {Foltz}, C.~B. 1992,
  \apj, 396, 487

\bibitem[{{Stockton} \& {MacKenty}(1987)}]{stoc87}
{Stockton}, A., \& {MacKenty}, J.~W. 1987, \apj, 316, 584

\bibitem[{{Sturm} {et~al.}(2011){Sturm}, {Gonz{\'a}lez-Alfonso}, {Veilleux},
  {Fischer}, {Graci{\'a}-Carpio}, {Hailey-Dunsheath}, {Contursi}, {Poglitsch},
  {Sternberg}, {Davies}, {Genzel}, {Lutz}, {Tacconi}, {Verma}, {Maiolino}, \&
  {de Jong}}]{stur11}
{Sturm}, E., {Gonz{\'a}lez-Alfonso}, E., {Veilleux}, S., {et~al.} 2011, \apjl,
  733, L16

\bibitem[{{Su} \& {Finkbeiner}(2012)}]{su12}
{Su}, M., \& {Finkbeiner}, D.~P. 2012, \apj, 753, 61

\bibitem[{{Su} {et~al.}(2010){Su}, {Slatyer}, \& {Finkbeiner}}]{su10}
{Su}, M., {Slatyer}, T.~R., \& {Finkbeiner}, D.~P. 2010, \apj, 724, 1044

\bibitem[{{Sun} {et~al.}(2014){Sun}, {Greene}, {Zakamska}, \&
  {Nesvadba}}]{sun14}
{Sun}, A.-L., {Greene}, J.~E., {Zakamska}, N.~L., \& {Nesvadba}, N.~P.~H. 2014,
  \apj

\bibitem[{{Tabor} \& {Binney}(1993)}]{tabo93}
{Tabor}, G., \& {Binney}, J. 1993, \mnras, 263, 323

\bibitem[{{Tielens}(2005)}]{tiel05}
{Tielens}, A.~G.~G.~M. 2005, {The Physics and Chemistry of the Interstellar
  Medium}

\bibitem[{{Treister} {et~al.}(2008){Treister}, {Krolik}, \&
  {Dullemond}}]{trei08}
{Treister}, E., {Krolik}, J.~H., \& {Dullemond}, C. 2008, \apj, 679, 140

\bibitem[{{Vanden Berk} {et~al.}(2001){Vanden Berk}, {Richards}, {Bauer},
  {Strauss}, {Schneider}, {Heckman}, {York}, {Hall}, {Fan}, {Knapp},
  {Anderson}, {Annis}, {Bahcall}, {Bernardi}, {Briggs}, {Brinkmann}, {Brunner},
  {Burles}, {Carey}, {Castander}, {Connolly}, {Crocker}, {Csabai}, {Doi},
  {Finkbeiner}, {Friedman}, {Frieman}, {Fukugita}, {Gunn}, {Hennessy},
  {Ivezi{\'c}}, {Kent}, {Kunszt}, {Lamb}, {Leger}, {Long}, {Loveday}, {Lupton},
  {Meiksin}, {Merelli}, {Munn}, {Newberg}, {Newcomb}, {Nichol}, {Owen}, {Pier},
  {Pope}, {Rockosi}, {Schlegel}, {Siegmund}, {Smee}, {Snir}, {Stoughton},
  {Stubbs}, {SubbaRao}, {Szalay}, {Szokoly}, {Tremonti}, {Uomoto}, {Waddell},
  {Yanny}, \& {Zheng}}]{vand01}
{Vanden Berk}, D.~E., {Richards}, G.~T., {Bauer}, A., {et~al.} 2001, \aj, 122,
  549

\bibitem[{{Veilleux}(1991{\natexlab{a}})}]{veil91a}
{Veilleux}, S. 1991{\natexlab{a}}, \apjs, 75, 357

\bibitem[{{Veilleux}(1991{\natexlab{b}})}]{veil91b}
---. 1991{\natexlab{b}}, \apjs, 75, 383

\bibitem[{{Veilleux}(1991{\natexlab{c}})}]{veil91c}
---. 1991{\natexlab{c}}, \apj, 369, 331

\bibitem[{{Veilleux} {et~al.}(2005){Veilleux}, {Cecil}, \&
  {Bland-Hawthorn}}]{veil05}
{Veilleux}, S., {Cecil}, G., \& {Bland-Hawthorn}, J. 2005, \araa, 43, 769

\bibitem[{{Veilleux} {et~al.}(1994){Veilleux}, {Cecil}, {Bland-Hawthorn},
  {Tully}, {Filippenko}, \& {Sargent}}]{veil94}
{Veilleux}, S., {Cecil}, G., {Bland-Hawthorn}, J., {et~al.} 1994, \apj, 433, 48

\bibitem[{{Veilleux} {et~al.}(1995){Veilleux}, {Kim}, {Sanders}, {Mazzarella},
  \& {Soifer}}]{veil95}
{Veilleux}, S., {Kim}, D.-C., {Sanders}, D.~B., {Mazzarella}, J.~M., \&
  {Soifer}, B.~T. 1995, \apjs, 98, 171

\bibitem[{{Veilleux} {et~al.}(2013){Veilleux}, {Mel{\'e}ndez}, {Sturm},
  {Gracia-Carpio}, {Fischer}, {Gonz{\'a}lez-Alfonso}, {Contursi}, {Lutz},
  {Poglitsch}, {Davies}, {Genzel}, {Tacconi}, {de Jong}, {Sternberg}, {Netzer},
  {Hailey-Dunsheath}, {Verma}, {Rupke}, {Maiolino}, {Teng}, \&
  {Polisensky}}]{veil13}
{Veilleux}, S., {Mel{\'e}ndez}, M., {Sturm}, E., {et~al.} 2013, \apj, 776, 27

\bibitem[{{Vignali} {et~al.}(2010){Vignali}, {Alexander}, {Gilli}, \&
  {Pozzi}}]{vign10}
{Vignali}, C., {Alexander}, D.~M., {Gilli}, R., \& {Pozzi}, F. 2010, \mnras,
  404, 48

\bibitem[{{Villar-Mart{\'{\i}}n} {et~al.}(2008){Villar-Mart{\'{\i}}n},
  {Humphrey}, {Mart{\'{\i}}nez-Sansigre}, {P{\'e}rez-Torres}, {Binette}, \&
  {Zhang}}]{vill08}
{Villar-Mart{\'{\i}}n}, M., {Humphrey}, A., {Mart{\'{\i}}nez-Sansigre}, A.,
  {et~al.} 2008, \mnras, 390, 218

\bibitem[{{Wagner} {et~al.}(2012){Wagner}, {Bicknell}, \& {Umemura}}]{wagn12}
{Wagner}, A.~Y., {Bicknell}, G.~V., \& {Umemura}, M. 2012, \apj, 757, 136

\bibitem[{{Wagner} {et~al.}(2013){Wagner}, {Umemura}, \& {Bicknell}}]{wagn13}
{Wagner}, A.~Y., {Umemura}, M., \& {Bicknell}, G.~V. 2013, \apjl, 763, L18

\bibitem[{{Wang} {et~al.}(2009){Wang}, {Fabbiano}, {Elvis}, {Risaliti},
  {Mazzarella}, {Howell}, \& {Lord}}]{wang09}
{Wang}, J., {Fabbiano}, G., {Elvis}, M., {et~al.} 2009, \apj, 694, 718

\bibitem[{{Weingartner} \& {Draine}(2001)}]{wein01}
{Weingartner}, J.~C., \& {Draine}, B.~T. 2001, \apj, 548, 296

\bibitem[{{Whittle}(1985{\natexlab{a}})}]{whit85a}
{Whittle}, M. 1985{\natexlab{a}}, \mnras, 213, 1

\bibitem[{{Whittle}(1985{\natexlab{b}})}]{whit85c}
---. 1985{\natexlab{b}}, \mnras, 216, 817

\bibitem[{{Whittle}(1992)}]{whit92b}
---. 1992, \apj, 387, 109

\bibitem[{{Wilson} \& {Heckman}(1985)}]{wils85}
{Wilson}, A.~S., \& {Heckman}, T.~M. 1985, in Astrophysics of Active Galaxies
  and Quasi-Stellar Objects, ed. J.~S. {Miller}, 39--109

\bibitem[{{Xu} {et~al.}(1999){Xu}, {Livio}, \& {Baum}}]{xu99}
{Xu}, C., {Livio}, M., \& {Baum}, S. 1999, \aj, 118, 1169

\bibitem[{{York} {et~al.}(2000)}]{york00}
{York}, D.~G., {et~al.} 2000, \aj, 120, 1579

\bibitem[{{Zakamska} {et~al.}(2008){Zakamska}, {G{\'o}mez}, {Strauss}, \&
  {Krolik}}]{zaka08}
{Zakamska}, N.~L., {G{\'o}mez}, L., {Strauss}, M.~A., \& {Krolik}, J.~H. 2008,
  \aj, 136, 1607

\bibitem[{{Zakamska} {et~al.}(2004){Zakamska}, {Strauss}, {Heckman},
  {Ivezi{\'c}}, \& {Krolik}}]{zaka04}
{Zakamska}, N.~L., {Strauss}, M.~A., {Heckman}, T.~M., {Ivezi{\'c}}, {\v Z}.,
  \& {Krolik}, J.~H. 2004, \aj, 128, 1002

\bibitem[{{Zakamska} {et~al.}(2003)}]{zaka03}
{Zakamska}, N.~L., {et~al.} 2003, \aj, 126, 2125

\bibitem[{{Zakamska} {et~al.}(2005)}]{zaka05}
---. 2005, \aj, 129, 1212

\bibitem[{{Zakamska} {et~al.}(2006)}]{zaka06}
---. 2006, \aj, 132, 1496

\bibitem[{{Zubovas} \& {King}(2012)}]{zubo12}
{Zubovas}, K., \& {King}, A. 2012, \apjl, 745, L34

\end{thebibliography}

\end{document}